\documentclass[journal]{IEEEtran}

\usepackage{mathrsfs}
\usepackage[noadjust]{cite}
\usepackage{graphicx,color,overpic,psfrag}
\usepackage{amsmath, amssymb}
\usepackage{latexsym}
\usepackage{bm}
\usepackage{amssymb}
\usepackage{cases}
\usepackage{array}
\usepackage{fancyhdr}
\usepackage{setspace}

\ifCLASSOPTIONcompsoc
\usepackage[caption=false,font=normalsize,labelfont=sf,textfont=sf]{subfig}
\else
\usepackage[caption=false,font=footnotesize]{subfig}
\fi

\usepackage{url}
\usepackage{algpseudocode}
\usepackage{algorithm}
\usepackage{blkarray}
\usepackage{booktabs}

\usepackage{multirow}
\usepackage{dsfont}
\usepackage{tabularx}
\usepackage[table]{xcolor}

\usepackage{amsfonts}

\usepackage{letltxmacro}

\graphicspath{{figure/}}

\newtheorem{remark}{Remark}

\newtheorem{prop}{Proposition}

\newcommand{\figref}[1]{Fig. \ref{#1}}

\newcommand{\alref}[1]{Algorithm \ref{#1}}
\newcommand{\appref}[1]{Appendix \ref{#1}}
\newcommand{\secref}[1]{Section \ref{#1}}

\newcommand{\propref}[1]{Proposition \ref{#1}}

\newcommand{\Exp}{{\mathsf{E}}}
\newcommand{\expect}[1]{\Exp\left\{#1\right\}}

\newcommand{\tr}[1]{\mathsf{tr}\left\{#1\right\}}
\newcommand{\diag}[1]{\mathsf{diag}\left\{#1\right\}}

\newcommand{\cK}{\mathcal{K}}
\newcommand{\cL}{\mathcal{L}}

\newcommand{\cO}{\mathcal{O}}
\newcommand{\cP}{\mathcal{P}}

\newcommand{\bn}{\mathbf{n}}

\newcommand{\bx}{\mathbf{x}}
\newcommand{\by}{\mathbf{y}}

\newcommand{\bB}{\mathbf{B}}

\newcommand{\bD}{\mathbf{D}}

\newcommand{\bG}{\mathbf{G}}
\newcommand{\bH}{\mathbf{H}}
\newcommand{\bI}{\mathbf{I}}

\newcommand{\bK}{\mathbf{K}}

\newcommand{\bQ}{\mathbf{Q}}
\newcommand{\bR}{\mathbf{R}}

\newcommand{\bU}{\mathbf{U}}
\newcommand{\bV}{\mathbf{V}}
\newcommand{\bW}{\mathbf{W}}
\newcommand{\bX}{\mathbf{X}}
\newcommand{\bY}{\mathbf{Y}}

\newcommand{\bzero}{\mathbf{0}}

\newcommand{\bLambda}{{\boldsymbol\Lambda}}
\newcommand{\bOmega}{{\boldsymbol\Omega}}
\newcommand{\bomega}{{\boldsymbol\omega}}

\newcommand{\upnot}[2]{#1^{\mathrm{#2}}}
\newcommand{\dnnot}[2]{#1_{\mathrm{#2}}}

\newcommand{\ntb}{\notag\\}

\newcommand{\figdoucolwid}{0.5\textwidth}

\newcommand{\R}{\mathbb{R}}
\newcommand{\C}{\mathbb{C}}
\newcommand{\Q}{\mathbf{Q}}

\newcommand{\I}{\mathbf{I}}
\newcommand{\de}{\mathrm{d}}
\newcommand{\K}{\cal{K}}

\newcommand{\Lda}{\mathbf{\Lambda}}
\newcommand{\Hk}{\mathbf{H}_{k}}

\newcommand{\GkH}{\mathbf{G}_{k}^{H}}
\newcommand{\Gk}{\mathbf{G}_{k}}

\newcommand{\kt}{{{\widetilde{\mathbf{K}}}_{k}}}
\newcommand{\Qk}{\mathbf{Q}_{k}}

\newcommand{\lambdak}{\mathbf{\Lambda}_{k}}

\newcommand{\SE}{\eta_{\mathrm{SE}}}
\newcommand{\EE}{\eta_{\mathrm{EE}}}
\newcommand{\RE}{\eta_{\mathrm{RE}}}
\newcommand{\Pc}{P_{\mathrm{c}}}
\newcommand{\Ps}{P_{\mathrm{s}}}

\newcommand{\Pmax}{P_{\mathrm{max}}}
\newcommand{\kbar}{{{\overline{\mathbf{K}}}_{k}}}
\newcommand{\Rbar}{{\overline{g}}}

\newcommand{\bit}{\mathrm {bits}}
\newcommand{\second}{\mathrm {s}}
\newcommand{\ub}{\mathrm {ub}}
\newcommand{\Hz}{\mathrm {Hz}}
\newcommand{\Jou}{\mathrm {Joule}}
\newcommand{\Ptot}{P_{\mathrm {tot}}}
\newcommand{\Psum}{P_{\mathrm {sum}}}
\newcommand{\PT}{P_{\mathrm {T}}}
\newcommand{\PTt}{\widetilde{P}_{\mathrm {T}}}
\newcommand{\Popt}{P_{\mathrm {opt}}^{(\ell)}}
\newcommand{\SEbar}{{\overline {\eta}}^{(\ell)}_{\mathrm{SE}}}
\newcommand{\REbar}{{\overline {\eta}}^{(\ell)}_{\mathrm{RE}}}
\newcommand{\EEbar}{{\overline {\eta}}^{(\ell)}_{\mathrm{EE}}}
\newcommand{\SEt}{{\widetilde {\eta}}_{\mathrm{SE}}}
\newcommand{\REt}{{\widetilde {\eta}}_{\mathrm{RE}}}
\newcommand{\EEt}{{\widetilde {\eta}}_{\mathrm{EE}}}

\newcommand{\Ta}{\triangle f_{k,\ub}^{(\ell)} \left(\Lda\right)}
\allowdisplaybreaks

\begin{document}

\title{Spectral Efficiency and Energy Efficiency Tradeoff in Massive MIMO Downlink Transmission\\ with Statistical CSIT}

\author{
Li~You, Jiayuan~Xiong,
Alessio~Zappone, Wenjin~Wang, and~Xiqi~Gao%
\thanks{
This work was presented in part at the IEEE Global Conference on Signal and Information Processing (GlobalSIP), Ottawa, ON, Canada, Nov. 2019 \cite{xiong2019energyspectral}.
}
\thanks{
L. You, J. Xiong, W. Wang, and X. Q. Gao are with the National Mobile Communications Research Laboratory, Southeast University, Nanjing 210096, China, and also with the Purple
Mountain Laboratories, Nanjing 211100, China (e-mail: liyou@seu.edu.cn; jyxiong@seu.edu.cn; wangwj@seu.edu.cn; xqgao@seu.edu.cn).
}
\thanks{
A. Zappone is with DIEI, University of Cassino and Southern Lazio, Via G. Di Biasio 43, 03043, Cassino, Italy. He is also with Consorzio Nazionale Interuniversitario per le Telecomunicazioni (CNIT), V.le G.P. Usberti  181/A, 43124, Parma, Italy (e-mail: alessio.zappone@unicas.it).
}
}

\maketitle

\begin{abstract}
As a key technology for future wireless networks, massive multiple-input multiple-output (MIMO) can significantly improve the energy efficiency (EE) and spectral efficiency (SE), and the performance is highly dependant on the degree of the available channel state information (CSI).
While most existing works on massive MIMO focused on the case where the instantaneous CSI at the transmitter (CSIT) is available, it is usually not an easy task to obtain precise instantaneous CSIT.
In this paper, we investigate EE-SE tradeoff in single-cell massive MIMO downlink transmission with statistical CSIT. To this end, we aim to optimize the system resource efficiency (RE), which is capable of striking an EE-SE balance. We first figure out a closed-form solution for the eigenvectors of the optimal transmit covariance matrices of different user terminals, which indicates that beam domain is in favor of performing RE optimal transmission in massive MIMO downlink. Based on this insight, the RE optimization precoding design is reduced to a real-valued power allocation problem. Exploiting the techniques of sequential optimization and random matrix theory, we further propose a low-complexity suboptimal two-layer water-filling-structured power allocation algorithm. Numerical results illustrate the effectiveness and near-optimal performance of the proposed statistical CSI aided RE optimization approach.
\end{abstract}

\begin{IEEEkeywords}
Energy efficiency, spectral efficiency, tradeoff, resource efficiency, massive MIMO, statistical CSI, power allocation.
\end{IEEEkeywords}
%

\section{Introduction}

\IEEEPARstart{D}{ue} to the sheer number of mobile devices and emerging applications, the demands of wireless data services have been drastically increasing in recent years. Those ubiquitous communication services require higher data transmission rates for massive connections and therefore pose new challenges for future wireless communications. Thanks to the deployment of large-scale antenna arrays at the base stations (BSs), massive multiple-input multiple-output (MIMO) can serve a large number of user terminals (UTs) over the same time/frequency resources \cite{Marzetta10Noncooperative}. Due to the significant potential gains in both energy efficiency (EE) and spectral efficiency (SE), massive MIMO has received extensive research interest and become an inevitable mainstream for next-generation wireless communications \cite{Sun2015Beam,bjornson2016massive}.

Energy aware optimization for wireless communications has received tremendous attention in the last few years, owing to both ecological and economical concerns \cite{zappone2015energy}.
Traditionally, SE is deemed to be a more critical design objective than EE to increase the transmission rate regardless of the cost. However, even though wireless networks may be able to achieve the required high data rates, power consumption might be dramatically increased, which accounts for the fundamentality and necessity of environment-friendly system designs. Consequently, green communication metrics such as EE have emerged as a vital design criterion for practical cellular networks \cite{Zappone2016energy}. However, EE-optimal strategies might sometimes collide with SE-optimal ones. Thus, how to strike a balance between EE and SE is worth investigation.

In the literature, extensive works have been emerging to cope with EE optimization wireless transmission design \cite{hellings2013energy,Jiang2011Downlink,Energy2012ShenZukang,Tervo2015Optimal,Energy2016QDVu}. For instance, an efficient algorithm aimed to reach the optimal EE for MIMO broadcast channels was proposed in \cite{hellings2013energy}. In \cite{Jiang2011Downlink}, the system EE for multiuser downlink was optimized with a zero-gradient-based iterative approach. In \cite{Energy2012ShenZukang}, a suboptimal solution to the multiuser EE optimization problem was obtained by upper-bounding the objective with a convex function. Low-complexity approaches were developed for joint design of energy efficient beamforming and antenna selection in \cite{Tervo2015Optimal}. Energy-efficient zero-forcing precoding strategy for small-cell networks was investigated in \cite{Energy2016QDVu}. Compared with the works focusing only on a single criterion, there are fewer works that investigate the EE-SE tradeoff, which can be generally classified into several categories. One is to jointly maximize EE and SE by introducing a tradeoff factor \cite{huang2018spectral,Energy2013HeChunlong,Energy2016LiuZujun}. The others are to maximize EE under a certain SE requirement \cite{Energy2011Xiong,Energy2012he,Energy2016Zhang} or vice versa \cite{Optimal2003Shen}. However, it is worth noticing that most existing works rely on the knowledge of instantaneous channel state information at the transmitter (CSIT). In practical systems, acquiring instantaneous CSIT is usually challenging, especially in the massive MIMO downlink. For instance, relying on channel reciprocity, downlink CSIT acquisition can be done via uplink training in time-division duplex (TDD) systems. However, the obtained downlink CSI may still be inaccurate due to practical limitations such as the calibration error in the radio frequency chains \cite{choi2014downlink}. Even worse, for frequency-division duplex (FDD) systems, acquiring downlink CSIT becomes more challenging due to the lack of channel reciprocity \cite{wang2016energy}. The feedback overhead for downlink CSI increases linearly with the number of transmit antennas when orthogonal pilot sequences are adopted, which might be prohibitive for practical massive MIMO systems \cite{You15Pilot,You16Channel}. Moreover, when the UTs are in high mobility, the obtained CSI quickly becomes outdated, especially when the feedback delay is larger than the channel coherence time. Compared with instantaneous CSI, the statistical CSI, e.g., the spatial correlation and the channel mean, is more likely to be stable during a longer period.
Note that it is in general not a difficult task for the BS to obtain the relatively slowly-varying statistical CSI through long-term feedback or covariance extrapolation \cite{wang2013precoder,wang2012statistical,khalilsarai2019fdd}. Therefore, exploiting statistical CSI for downlink precoder design is a promising approach in practical massive MIMO systems.

To this end, we consider EE-SE tradeoff design for massive MIMO downlink precoding with statistical CSIT. We adopt a flexible and unified green metric, namely, resource efficiency (RE) \cite{tang2014resource} and investigate RE optimization to acquire an adaptive EE-SE tradeoff.
The main contributions of this paper are summarized as follows:
\begin{itemize}
\item We investigate the transmission strategy for RE maximization in massive MIMO downlink. The alternating optimization approach is adopted to handle this complicated matrix-valued optimization problem with numerous variables. By first deriving a necessary condition that the optimal transmit covariance matrices should follow, we show that as the number of transmit antennas grows to infinity, the eigenvectors of the optimal transmit covariance matrices of different UTs asymptotically become identical. As a consequence, beam domain transmission becomes favorable for statistical CSI aided resource efficient massive MIMO downlink transmission.
\item Guided by the above insight, we reduce the original problem to a power allocation problem, which aims to maximize the system RE for massive MIMO downlink in the beam domain. We derive a deterministic equivalent (DE) of the system RE to simplify the computational complexity of the power allocation problem. Later, exploiting the minorization-maximization (MM) algorithm, the RE maximization problem with non-convex fractional objectives is then converted into a series of strictly quasi-concave optimization subproblems.
\item Utilizing the inherent properties of the strictly quasi-concave subproblems, we decompose them into two layers, where we aim to find the transmit power and the corresponding power allocation matrices, respectively. Based on the two-layer decomposition, we develop a well-structured suboptimal algorithm with guaranteed convergence, where a derivative-assisted gradient approach and a water-filling scheme are applied in the outer- and inner-layer optimization, respectively. Numerical results illustrate the near-optimality of the proposed RE maximization iterative algorithm to obtain a reasonable and adjustable EE-SE tradeoff.
\end{itemize}

The rest of the paper is organized as follows. In \secref{sec:system model}, we introduce the system model as well as the definition of system RE. In \secref{sec:opt_RE_design}, we investigate the transmission strategy design for RE maximization with statistical CSIT. We first show that beam domain transmission is favorable for RE optimization. Then we develop a suboptimal low-complexity and well-structured power allocation algorithm for RE maximization. The simulation results are drawn in \secref{sec:numerical_results}. The conclusion is presented in \secref{sec:conclusion}.

We adopt the following notations throughout the paper. Upper-case bold-face letters denote matrices, and lower-case bold-face letters denote column vectors, respectively. We use ${\mathbf{I}}_M$ to denote the $M \times M$ identity matrix where the subscript is omitted when no confusion caused. The superscripts ${( \cdot )^{ - 1}}$, ${( \cdot )^T}$, and ${( \cdot )^H}$ represent the matrix inverse, transpose, and conjugate-transpose operations, respectively. The ensemble expectation, matrix trace, and determinant operations are represented by $\expect{ \cdot }$, $\tr{\cdot }$, and $\mathrm{det} ( \cdot )$, respectively. The operator $\diag{\bf{x}}$ indicates a diagonal matrix with $\bf{x}$ along its main diagonal. We use ${[{\bf{A}}]_{m,n}}$ to represent the $(m,n)$th element of matrix ${\bf{A}}$. The inequality ${\bf{A}} \succeq \bzero$ means that ${\bf{A}}$ is Hermitian positive semi-definite. The notation $[x]^+$ denotes $\max(x, 0)$. The operator $\odot$ denotes the Hadamard product. The notation $\triangleq$ is used for definitions.

\section{System Model}\label{sec:system model}

\subsection{Channel Model}
Consider a single-cell massive MIMO downlink where one BS simultaneously transmits signals to a set of $K$ multiple-antenna UTs, which is denoted by $\cK \triangleq \left\{ {1,2, \ldots ,K} \right\}$. The BS has $M$ antennas and each UT $k \in \cK $ has $N_k$ receive antennas.
The MIMO channel spatial correlations are described by the jointly correlated Rayleigh fading model \cite{Gao09Statistical}, where the downlink channel matrix $\Hk \in {\C ^{N_k \times M}}$ from the BS to UT $k$ follows the structure as
\begin{align}\label{eq:beam_H}
\bH_k = \bU_k \bG_k \bV_k^H
\end{align}
where $\bU_k \in {\C ^{N_k \times N_k}}$ and $\bV_k \in {\C ^{M\times M}}$ are both deterministic unitary matrices, representing the eigenvectors of the receive correlation matrix and the BS correlation matrix of $\bH_{k}$, respectively.
For our considered massive MIMO channels, when $M \to \infty $, $\bH_k$ can be well approximated by \cite{You15Pilot,You16Channel}
\begin{align}\label{eq:Hk}
\bH_k \mathop  = \limits^{M \to \infty } \bU_k \bG_k \bV^H.
\end{align}
It has been widely recognized that ${\bf{V}}$ is independent of the locations of UTs and only depends on the BS antenna array geometry in massive MIMO \cite{Adhikary13Joint}. For example, for the uniform linear array (ULA) case with antenna spacing of half-wavelength, the discrete Fourier transform (DFT) matrix can be used to well approximate $\bV$ \cite{You15Pilot,You16Channel}.
Besides, $\bG_k \in {\C ^{N_k \times M}}$ in \eqref{eq:beam_H} is referred to as the beam domain channel matrix \cite{You17BDMA}, whose elements are zero-mean and independently distributed.
The statistical CSI of $\bG_k$, i.e., the eigenmode channel coupling matrix \cite{Gao09Statistical}, is modeled as
\begin{align}
\bOmega_k = \expect { \bG_k \odot \bG_k^* }  \in { \R ^{N_k \times M}}.
\end{align}
Assume that the BS has the knowledge of statistical CSI, i.e., $\bOmega_k(\forall k)$, via channel sounding \cite{Lu2019Robust}. Denoted by $\bx \in {\C^{M \times 1}}$ the input of the massive MIMO downlink transmission, the received signal of UT $k$ is
\begin{align}
\by_k = \bH_k \bx + \bn_k \in {\C ^{N_k \times 1}}
\end{align}
where $\bn_k \in {{\mathbb{C}}^{N_k\times 1}}$ denotes the circularly symmetric complex Gaussian noise with zero mean and covariance matrix ${\sigma ^2}{\bI_{N_k}}$. Note that $\bx = \sum\nolimits_{k} {\bx_k}$, where $\bx_k$ is the signal vector intended for UT $k$ with $\bQ_k = \expect { \bx_k \bx_k^H} \in {\C ^{M \times M}}$ being its covariance matrix. In addition, the transmit signal $\bx_k(\forall k)$ satisfies $ \expect { \bx_k}  = \bzero$ and $ \expect { \bx_k \bx_{k'}^H}  = \bzero \left( {\forall k' \ne k} \right)$.

\subsection{System SE}

Assume that each UT $k$ has access to instantaneous CSI of its own channel with properly designed pilot signals \cite{Sun2015Beam}. From a worst-case design perspective \cite{Hassibi2003How}, the aggregate interference-plus-noise $\bn_k' = \sum\nolimits_{i \ne k} {\bH_k \bx_i}  + \bn_k$ at UT $k$ is treated as the Gaussian noise. Then, the following ergodic rate\footnote{Note that the ergodic rate can be approached via using a long coding length over fast fading channels with a large delay \cite{Tse05Fundamentals}.} of UT $k$ can be achievable
\begin{align}\label{eq:ergodic_rate1}
\upnot{R_k}{ach} = \expect{\log\det\left(\bI_{N_k}+\bH_k\bQ_k\bH_k^H\bW_k^{-1}\right)}
\end{align}
where $\bW_{k}$ is covariance of $\bn_k'$ given by
\begin{align}
\bW_k = \sigma^2\bI_{N_k} + \sum\nolimits_{i\ne k}^{K}{\bH_k\bQ_i\bH_k^H}.
\end{align}
Motivated by the channel hardening effect in massive MIMO, we approximate $\bW_k$ by its expectation over $\bH_k$ as
\begin{align}
\bW_k \approx \bK_k &= \expect{\bW_k} \ntb
&= {\sigma ^2} \bI_{N_k} + \sum\nolimits_{i\ne k}^{K} {\expect { \bH_k \bQ_i \bH_k^H } }.
\end{align}
Then, an approximated ergodic data rate of UT $k$ can be expressed by \cite{Lu2019Robust,wu18beam,You2020energy}
\begin{align}\label{eq:ergodic_rate}
R_k & = \expect{\log\det\left(\bI_{N_k}+\bH_k\bQ_k\bH_k^H\bK_k^{-1}\right)} \ntb
& = \expect { \log \det \left( \bK_k + \bH_k \bQ_k \bH_k^H \right) }  -  \log\det \left( \bK_k \right)\ntb
& =  \expect { \log\det \left( \kt + \bG_k \bV^H \bQ_k \bV \bG_k^H \right) }  -  \log\det\left( \kt \right)
\end{align}%
where the third equality above follows from rewriting $\bH_k$ with the aid of \eqref{eq:Hk} and applying Sylvester's determinant identity, i.e., $\det \left( \bI + \bX \bY \right) = \det \left( \bI + \bY \bX \right)$. In addition, $\kt$ in \eqref{eq:ergodic_rate} is defined as
\begin{align}\label{eq:K_tilde}
{\kt} &\triangleq \bU_k^H \bK_k \bU_k \ntb
& = {\sigma ^2} \bI_{N_k} + \sum\nolimits_{i \ne k}^K {\underbrace {\expect { \bG_k \bV^H \bQ_i \bV \bG_k^H } }_{ \triangleq {{\bf{\Pi}}_k}\left( \bV^H \bQ_i \bV \right)}}  \in { \C ^{N_k \times N_k}}.
\end{align}

Note that ${{\bf{\Pi}}_k}\left( \bX \right)$ defined in \eqref{eq:K_tilde} is a matrix-valued function of $\bX$. Utilizing the independently distributed properties of the elements of the beam domain channel $\Gk$, it can be shown that ${{\bf{\Pi}}_k}\left( \bX \right)$ is diagonal with the diagonal elements given by
\begin{align}\label{eq:A}
{\left[ {{{\bf{\Pi}}_k}\left( \bX \right)} \right]_{n,n}}
= \tr { \diag{\left( {\left[ \bOmega_k \right]}_{n,:}\right)^T}  \bX }.
\end{align}
Since the rate expression in \eqref{eq:ergodic_rate} is shown to be a good approximation in the numerical results, we will use this approximated rate expression in the rest of the paper.
Then, the system SE can be defined as the sum rate of all UTs given by
\begin{align}\label{eq:SE_definition}
\SE = \sum\nolimits_{k = 1}^K {{R_k}} \  ( \bit / \second / \Hz ).
\end{align}

\subsection{System EE}
To describe the EE metric, we first introduce an affine power consumption model \cite{xu2013energy}, where the overall power consumption is comprised of three parts, i.e.,
\begin{align}\label{eq:power_consumption_model}
\Psum = \xi \sum\nolimits_{k=1}^{K}{\tr { \Qk } } + M{\Pc} +\Ps
\end{align}
where the scaling coefficient $\xi$ describes the transmit amplifier inefficiency, $ \sum\nolimits_{k}{ \tr {\Qk} }$ represents the overall transmit power, $\Pc$ denotes the dynamic power dissipations per antenna (e.g., power consumption in the digital-to-analog converter, the frequency synthesizer, and the BS filter and mixer), which is independent of $\sum\nolimits_k {\tr { \Qk } }$, and $\Ps$ incorporates the static circuit power consumption, which is independent of both $M$ and $\sum\nolimits_k {\tr { \Qk } }$.\footnote{Note that the power consumption parameters, such as $\Pc$ and $\Ps$, in \eqref{eq:power_consumption_model} are also related to the system bandwidth in practice \cite{Mezghani2010circuit}. In our optimization, we assume a predefined fixed system bandwidth, and thus the adopted power consumption model in \eqref{eq:power_consumption_model} makes sense.}

In practical systems, the BS overall transmit power is usually limited, leading to the following constraint
\begin{align}\label{eq:constraint1}
 \sum\nolimits_{k=1}^{K}{\tr { \Qk }}\le P_{\mathrm{\max }}
\end{align}
where $P_{\mathrm{\max }}$ is related to the BS transmit power budget. Under the above modeling of the ergodic rate in \eqref{eq:ergodic_rate} and the power consumption in \eqref{eq:power_consumption_model}, we define the system EE as follows
\begin{align}\label{eq:EE_definition}
\EE = W\frac{\SE  }{\Psum} \ \ ( \bit / \Jou )
\end{align}
where $W$ represents the system bandwidth.

\subsection{Problem Formulation}
Since both SE and EE are important metrics for communication system design, how to strike a balance between them is worth investigating. To this end, we consider the optimization of EE and SE simultaneously to obtain an EE-SE tradeoff by means of maximizing a weighted sum of EE and SE. However, directly adding SE and EE seems to be inappropriate due to the inconsistency between the metric units of SE and EE, which are bits/s/Hz and bits/Joule, respectively. Hence, we adopt a system design metric referred to as RE \cite{tang2014resource} given by
\begin{align}\label{eq:RE_definition}
\RE \triangleq \frac{\EE}{W} + \beta \frac{{\SE}}{{\Ptot}} \ \ ( \bit / \Jou / \Hz ) \quad
\end{align}
where $\beta(>0)$ is a weighting factor. Notice that the RE metric does have the ability to achieve an EE-SE tradeoff with $\beta$ in control of the EE-SE balance. In addition, $ \frac{1} {W}$ and $ \frac{1} {\Ptot}$ are both unit normalizer, where $\Ptot$ represents the BS overall power budget as
\begin{align}\label{eq:ptot}
\Ptot = \xi \Pmax + M{\Pc} + \Ps
\end{align}
which is similar to the power consumption model in \eqref{eq:power_consumption_model}. Moreover, let $\beta\frac{ W}{\Ptot}\triangleq \alpha/\left(1-\alpha\right)$, we can observe that maximizing $\RE$ is equivalent to maximizing $\left(1-\alpha\right)\EE + \alpha\SE$. Thus, maximizing the RE is equivalent to obtaining the Pareto optimum of the EE-SE multi-objective optimization problem \cite{Boyd04Convex}.

In the following, we investigate the precoding strategy design for massive MIMO downlink transmission under the RE maximization criterion, which is formulated as
\begin{align}\label{eq:problem1}
\cP_1:\quad\underset{{{\Q}_{1}},{{\Q}_{2}},\ldots ,{{\Q}_{K}}} \max  &\quad \RE \ntb
{\mathrm{s.t.}}\quad
&\quad \sum\nolimits_{k=1}^{K}{\tr { \Qk } }\le P_{\mathrm{\max }}\ntb
&\quad \Qk \succeq \bzero,\quad \forall k\in {\cal K}.
\end{align}
\begin{remark}
Different EE-SE tradeoff strategies can be achieved through changing the weighting factor $\beta$, which is decided by the system designers. For instance, the objective of problem $\cP_1$ degenerates to the system EE (with bandwidth normalization) when $\beta = 0$ and reduces to the system SE when $\beta \to \infty $.
\end{remark}

\section{Transmission Design for RE Maximization}\label{sec:opt_RE_design}

In this section, we study the transmission strategy for the RE maximization problem $\cP_1$ in \eqref{eq:problem1}.
The challenges in addressing problem $\cP_1$ lie in several aspects. First of all, the number of optimization variables in the matrix-valued problem $\cP_1$ is $M^2K$, which can be quite large since $M$ is usually large in practical massive MIMO systems. Secondly, the expectation operation involved in calculating the objective function of $\cP_1$ usually requires stochastic programming approaches, which will yield huge computational complexity. Moreover, the objective of $\cP_1$ involves a fractional function, $\EE$, with its numerator being non-concave, which adds more difficulty in solving $\cP_1$. In the following, we aim to develop an efficient suboptimal approach to address this challenging problem.

To solve the RE maximization problem $\cP_1$ more conveniently, we first decompose the transmit covariance matrix of UT $k$ as ${{\bf{Q}}_k} = {{\bf{\Psi }}_k}{\bLambda_k}{\bf{\Psi }}_k^H$ by eigenvalue decomposition. Note that the columns of ${\bf{\Psi }}_k$ and the corresponding diagonal elements in $\bLambda_k$ are the eigenvectors and the eigenvalues of $\bQ_{k}$, respectively. In fact, the eigenmatrix (the matrix consisting of all eigenvectors) ${\bf{\Psi }}_k$ represents the subspace where the transmit signal lies in. Moreover, the elements of the diagonal matrix $\bLambda_k$ represent the power assigned to each dimension/direction of the subspace for the transmit signals.

\subsection{Optimal Transmission Direction}

First, we investigate the optimal transmit signal directions of all UTs. Taking advantage of the massive MIMO channel characteristics, we identify the optimal eigenmatrix ${\bf{\Psi }}_k (\forall k)$ of the transmit signal covariance ${{\bf{Q}}_k} (\forall k)$ in the following proposition.

\begin{prop}\label{theorem:beam_domain_optimal}
The eigenvectors of the optimal transmit covariance matrices $\Qk$ for all $k$ to problem \eqref{eq:problem1} are all given by the columns of $\bV$, i.e.,
\begin{align}
\bQ_k^{\mathrm{opt}} = \bV \bLambda_k \bV^H,\forall k.
\end{align}
\end{prop}

\begin{IEEEproof}
See \appref{app:A}.
\end{IEEEproof}

\propref{theorem:beam_domain_optimal} reveals that to maximize the system RE in problem $\cP_1$, the optimal directions for the downlink transmit signals should be aligned with the eigenvectors of the BS correlation matrices, thereby being (asymptotically) independent of UTs. Thus, if we transform the signals into the beam domain, the condition in \propref{theorem:beam_domain_optimal} can be satisfied. In other words, beam domain transmission is favorable for RE optimization in downlink massive MIMO.

With a slight abuse of notations, we denote by $\Lda \triangleq \left\{ {{\Lda}_{1}},\ldots ,{{\Lda}_{K}} \right\}$. Then, according to \propref{theorem:beam_domain_optimal}, we can reduce the RE optimization problem over $\bQ_k  (\forall k)$ to the problem over $\bLambda$ as follows
\begin{align}\label{eq:problem2}
\cP_2:\quad\underset{\bLambda} \max \quad & \REt(\Lda) = \frac{\EEt(\Lda)}{W} + \beta \frac{{\SEt }(\Lda)}{{\Ptot}}\ntb
{\mathrm{s.t.}}\quad
& \sum\limits_{k=1}^{K}{\tr{ \lambdak }} \le \Pmax \ntb
& \lambdak \succeq \bzero,\; \lambdak\; \mathrm{diagonal},\; \forall k\in \K
\end{align}
where
\begin{align}
\EEt (\Lda) & = \frac{W\SEt \left(\Lda\right) }{\xi \sum\nolimits_{k=1}^{K}{\tr { \lambdak } } + M{\Pc} +\Ps} \\ \label{eq:SEt}
\SEt (\Lda) & = \sum\nolimits_{k = 1}^K \left\lbrace g_k\left(\bLambda\right) - f_k\left(\bLambda\right)\right\rbrace\\
g_k\left(\bLambda\right) &\triangleq \expect { \log \det \left( \kbar\left(\Lda\right)+ \Gk\lambdak\GkH \right)  }\\
f_k\left(\bLambda\right) &\triangleq \log \det \left( \kbar\left(\Lda\right) \right) \label{eq:fkla} \\
\kbar\left(\Lda\right) & \triangleq {\sigma ^2} \I_{N_k} +  \sum\nolimits_{i\ne k}^{K}{{{\bf{\Pi}}_k}({\bLambda_i})}.
\end{align}

With the above formulation, problem $\cP_2$ now turns out to be a power allocation problem in the beam domain.
Since the power allocation matrices $\bLambda_k (\forall k)$ are diagonal, the number of optimization variables is reduced from $M^2K$ in the original matrix-valued problem $\cP_1$ to $MK$ in problem $\cP_2$. In addition, $\bQ_k$ is complex-valued while $\bLambda_k$ is real-valued. Therefore, $\cP_2$ is a much simpler power allocation problem compared with the original precoding design problem $\cP_1$.

\subsection{Deterministic Equivalent Method}
Before solving $\cP_2$, we first introduce the DE approach to further reduce the optimization complexity. It is worth mentioning that while calculating the objective in $\cP_2$ in each iteration, manipulating the expectation operation through Monte-Carlo method is quite computationally cumbersome. Note that the DE method can provide deterministic approximations of the random matrix functions and the approximations are (almost surely) asymptotically accurate as the matrix sizes tend to infinity \cite{Couillet11Random,Lu16Free}. Therefore, we replace the rate expression by its DE to avoid channel averaging required in Monte-Carlo method. Specifically, the DE of $g_{k}\left( \Lda \right)$ is computed by
\begin{align}\label{eq:DE}
\Rbar_{k} \left( \Lda \right) &= \log \det \left( \mathbf{I}_{M}+\mathbf{\Gamma }_{k}{{\mathbf{\Lambda }}_{k}} \right)
+ \log \det \left( \mathbf{\widetilde{\Gamma }}_{k}+\mathbf{\overline{K}}_{k}\left( \Lda \right) \right) \ntb
&\qquad -\tr { \mathbf{I}_{N_k}-{{\mathbf{\widetilde{\Phi }}}^{-1}_{k}} }.
\end{align}
In \eqref{eq:DE}, $\mathbf{\Gamma }_{k}$ and $\mathbf{\widetilde{\Gamma }}_{k}$ are given by
\begin{align}
\label{eq:gamma_1}
\mathbf{\Gamma }_{k}&={{\bf{\Xi}}_{k}}\left( {{{\mathbf{\widetilde{\Phi }}}^{-1}_{k}}}{ {\left(\mathbf{\overline{K}}_{k}\left( \Lda \right)\right)} ^{-1}} \right)\\
\label{eq:gamma_2}
\mathbf{\widetilde{\Gamma }}_{k}&={{\bf{\Pi}}_{k}}\left( { {{{\mathbf{\Phi }}^{-1}_{k}}}}{{ {\mathbf{\Lambda }}_{k}} } \right)
\end{align}
respectively, where the DE auxiliary variables $\mathbf{\widetilde{\Phi }}_{k}$ and $\mathbf{\Phi }_{k}$ can be obtained via using the following iterative equations
\begin{align}
\label{eq:phi_1}
\mathbf{\widetilde{\Phi }}_{k}&=\mathbf{I}_{N_k}+{{\bf{\Pi}}_{k}}\left( { {{{\mathbf{\Phi }^{-1}_{k}}}}}{{\mathbf{\Lambda }}_{k}} \right){{\left(\mathbf{\overline{K}}_{k}\left( \Lda \right)\right)}^{-1}}\\
\label{eq:phi_2}
\mathbf{\Phi }_{k}&=\mathbf{I}_{M}+{{\bf{\Xi }}_{k}}\left( { {{{\mathbf{\widetilde{\Phi }}}^{-1}_{k}}}}{ {\left(\mathbf{\overline{K}}_{k}\left( \Lda \right)\right)} ^{-1}} \right){{\mathbf{\Lambda }}_{k}}.
\end{align}
The above matrix-valued function ${{\bf{\Xi}}_k}  \left( {\bf{X}} \right)$ is defined as
\begin{align}
\label{eq:Xi}
{{\bf{\Xi}}_k}  \left( {\bf{X}} \right) \triangleq \expect { {\bf{G}}_k^H {\bf{X}} \Gk }.
\end{align}
Exploiting the independently distributed properties of the beam domain channel elements, we can show that ${{\bf{\Xi}}_k}  \left( {\bf{X}} \right)$ is diagonal and its diagonal elements are represented by
\begin{align}\label{eq:Xi_element}
{\left[ {{{\bf{\Xi}}_k}({\bf{X}})} \right]_{m,m}} = \tr { \diag{{\left[{\bf{\Omega}}_k \right]}_{:,m}} {\mathbf{X}} }.
\end{align}
Consequently, we can obtain that $\mathbf{\widetilde{\Phi }}_{k}$, $\mathbf{\Phi }_{k}$, $\mathbf{\Gamma }_{k}$, and $\mathbf{\widetilde{\Gamma }}_{k}$ are all diagonal and the DE expression $\overline{g}_k\left(\bLambda\right)$ can be efficiently calculated. Then, with the aid of \eqref{eq:DE}, we turn to consider the following optimization subproblems
\begin{align}\label{eq:problemDE1}
\cP_3:\quad \underset{\Lda} \max \quad& \left( {\frac{1}{{\xi \sum\nolimits_{k}{\tr { \lambdak } } + M\Pc + \Ps}} + \frac{\beta}{{\Ptot}}} \right)  \ntb
&\qquad\cdot\left[{\sum\nolimits_{k=1}^{K}{ \left( \Rbar_{k}\left(\Lda\right) - f_{k}\left(\Lda\right) \right) }}\right]  \ntb
{\mathrm{s.t.}}\quad
& \sum\limits_{k=1}^{K}{\tr{ \lambdak }} \le \Pmax \ntb
& \lambdak \succeq \bzero,\; \lambdak\; \mathrm{diagonal},\; \forall k\in \K
\end{align}
where $f_k\left(\bLambda\right) $ is defined in \eqref{eq:fkla}.

From \eqref{eq:DE}, we observe that the DE expression $\Rbar_{k} \left( \Lda \right)$ depends mainly on ${{\bf{\Xi}}_{k}}({\bf{X}})$ and ${{\bf{\Pi}}_{k}}({{\bf{X}}})$, which are both computationally efficient. Consequently, the replacement of $g_{k} \left( \Lda \right)$ with $\Rbar_{k} \left( \Lda \right)$ leads to lower computational complexity in solving $\cP_3$ compared with $\cP_2$. In addition, the optimal solution to $\cP_3$ is an asymptotically optimal solution to $\cP_2$. Note that the (strict) concavity of $\Rbar_{k} \left( \Lda \right)$ over $\Lda$ can be obtained from \cite{Dumont2010On,Dupuy2011On}. Although utilizing the DE expression is more computationally efficient than Monte-Carlo method, solving problem $\cP_3$ is still challenging, due to the fact that the objective of $\cP_3$ is not concave in general. In the following, we proceed to solve subproblem $\cP_3$ for resource efficient beam domain power allocation.

\subsection{MM-based Power Allocation Algorithm}

Note that the objective of $\cP_3$ in \eqref{eq:problemDE1} involves a fractional function with respect to $\Lda$. Besides, $\Rbar_{k}\left( \Lda \right)$ and $f_{k} \left( \Lda \right)$ are both concave over $\Lda$, leading to a non-concave numerator of the fractional term, $\sum\nolimits_{k=1}^{K}{ \left( \Rbar_{k}\left(\Lda\right) - f_{k}\left(\Lda\right) \right) }$, in the objective of $\cP_3$. Therefore, directly utilizing classical fractional programming approaches might exhibit an exponential complexity \cite{zappone2015energy}. This calls for the development of a low-complexity approach for the considered beam domain power allocation problem.
In the following, we develop an efficient power allocation approach for RE maximization by means of sequential convex optimization tools. Specifically, we resort to the MM algorithm  \cite{sun2017majorization} to handle $\cP_3$ and the main idea of the MM algorithm lies in converting a non-convex problem to a series of easy-to-handle subproblems. From $\cP_3$, we can find that the numerator of the fractional term in the objective, $\sum\nolimits_{k=1}^{K}{ \left( \Rbar_{k}\left(\Lda\right) - f_{k}\left(\Lda\right) \right) }$, is the difference between two concave functions. Denoting by $\triangle f_{k,\ub}(\Lda)$ the first-order Taylor expansion of the negative rate term $f_{k}(\Lda)$, we have $f_k(\Lda) \le \triangle f_{k,\ub}(\Lda) $. Then, replacing $f_{k}(\Lda)$ with its first-order Taylor expansion $\triangle f_{k,\ub}(\Lda)$, the non-concave term $\sum\nolimits_{k=1}^{K}{ \left( \Rbar_{k}\left(\Lda\right) - f_{k}\left(\Lda\right) \right) }$ in problem $\cP_3$ can be lower-bounded by a concave function. This approximation has been used in some previous works \cite{zappone16energy,sun2017bdma,wu18beam}, where its effectiveness has also been verified. By doing so, problem $\cP_3$ is tackled through solving the following optimization subproblems
\begin{align}\label{eq:problem5}
\cP^{(\ell)}_4:\; \underset{\Lda} \max \quad& \left( {\frac{1}{{\xi \sum\nolimits_{k}{\tr { \lambdak } } + M\Pc + \Ps}} + \frac{\beta}{{\Ptot}}} \right)  \ntb
&\qquad\cdot\left[{\sum\nolimits_{k=1}^{K}{ \left( \Rbar_{k}\left(\Lda\right) - \triangle f_{k,\ub}^{(\ell)}\left(\Lda\right)\right) }} \right] \ntb
{\mathrm{s.t.}}\quad
& \sum\limits_{k=1}^{K}{\tr{ \lambdak }} \le \Pmax \ntb
& \lambdak \succeq \bzero,\; \lambdak\; \mathrm{diagonal},\; \forall k\in \K
\end{align}
where
\begin{align}
&\triangle f_{k,\ub}^{(\ell)} \left(\Lda\right) =  f_{k}\left(\Lda^{(\ell)}\right) \ntb
&\quad + \tr { {\left( {\frac{\partial }{{\partial {\bLambda_k}}}\sum\nolimits_{k'} {f_{k'}\left(\Lda^{(\ell)}\right) } } \right)}^{T} \left( \lambdak - {\lambdak^{(\ell)}} \right)}
\end{align}
where $\Lda^{\left( \ell \right)} \triangleq \left\{ {{\Lda}^{(\ell)}_{1}},{{\Lda}^{(\ell)}_{2}},\ldots ,{{\Lda}^{(\ell)}_{K}} \right\}$ and $\ell$ denotes the iteration index.
Moreover, the derivative $\frac{\partial }{{\partial {\bLambda_k}}}\sum\nolimits_{k'=1}^K {f_{k'}\left(\Lda^{(\ell)}\right) }$ can be derived as
\begin{align}\label{eq:derivative_CCCP}
{{\bf{D }}^{(\ell)}_{k}}  &\triangleq \frac{\partial }{{\partial {\bLambda_k}}}\sum\nolimits_{k'=1}^K {f_{k'}\left(\Lda^{(\ell)}\right) } \ntb
&= \sum\nolimits_{k' \ne k} {\sum\nolimits_{n = 1}^{N_{k'}} {\frac{{{{\widehat {\bf R}}_{k',n}}}}{{ {\sigma}^2 + \tr {{\bLambda^{(\ell)}_{{\backslash k'}}}{{\widehat {\bf R}}_{k',n}}}}}} }
\end{align}
where ${{\bLambda^{(\ell)}_{{\backslash k'}}}}= \sum\nolimits_{i \ne k'} {{\bLambda^{(\ell)}_{i}}}$ and ${{\widehat {\bf R}}_{k',n}} = \diag{ {\bomega } _{k',n}}$ with ${{\bomega } _{k',n}^{T}}$ being the $n$th row of ${{\bf{\Omega }}_{k'}}$. Note that ${{\bf{D }}^{(\ell)}_{k}}$ is a diagonal matrix with the corresponding $t$th diagonal entry given by
\begin{align}
&{[{{\bf{D }}^{(\ell)}_{k}}]_{t,t}} \ntb
& =\sum\nolimits_{k' \ne k} {\sum\nolimits_{n = 1}^{N_{k'}} {\frac{{{{[{{\bf{\Omega }}_{k'}}]}_{n,t}}}}{{{\sigma}^2 +  \sum\nolimits_{i \ne k'}^K {\sum\nolimits_{m = 1}^M {{{[{\bLambda^{(\ell)}_{i}}]}_{m,m}}{{[{{\bf{\Omega }}_{k'}}]}_{n,m}}} } }}} }.
\end{align}

\begin{prop}\label{prop:mm_convergence}
The objective value sequence of $\left\{\cP_4^{(\ell)}\right\}_{\ell=0}^{ \infty }$ is non-decreasing and guaranteed to converge.
In addition, every limit point of the power allocation sequence to $\left\{\cP_4^{(\ell)}\right\}_{\ell=0}^{ \infty }$ is a Karush-Kuhn-Tucker (KKT) point of problem $\cP_3$.
Moreover, upon the convergence of the objective value sequence of $\left\{\cP_4^{(\ell)}\right\}_{\ell=0}^{ \infty }$, the resulting power allocation solution satisfies the KKT optimality conditions of problem $\cP_3$.
\end{prop}
\begin{IEEEproof}
See \appref{app:B}.
\end{IEEEproof}

\subsection{Derivative-Assisted Gradient Approach}
Basically, $\cP^{(\ell)}_4$ in \eqref{eq:problem5} is still challenging to obtain the optimal solution. Unlike the SE maximization problem, transmission with full power budget might lead to reduced EE owing to the fact that EE will saturate when the excessive total power is consumed, and thus might not be optimal for EE-SE tradeoff. Therefore, seeking the transmit power consumption is critical to RE optimization. Motivated by this insight, we tackle $\cP^{(\ell)}_4$ through solving two nested problems, one for the allocation of the transmit power across the beams, and the other for the optimization of the transmit power. Specifically, the first one for power allocation is characterized as
\begin{align}\label{eq:ratemax}
\cP^{(\ell)}_5:\quad \SEbar \left( {\PT} \right) =  \max \limits_{\Lda} \quad & {\sum\nolimits_{k=1}^{K}{ \left( \Rbar_{k}\left(\Lda\right) - \Ta \right) }}\ntb
{\mathrm{s.t.}}\quad
& \lambdak \succeq \bzero,\; \lambdak\; \mathrm{diagonal},\; \forall k\in \K \ntb
& \sum\nolimits_{k=1}^{K}{\tr { \lambdak }} = \PT
\end{align}
where we introduce an auxiliary power variable $P_{\rm{T}}$ representing the overall transmit power and an auxiliary function $\SEbar \left( {\PT} \right)$, which is the maximal objective value of $\cP^{(\ell)}_5$. Note that $\SEbar \left( {\PT} \right)$ is the corresponding maximal system SE in the $\ell$th iteration of the MM method with a given $P_{\rm{T}}$. Then, we consider the other problem for the optimization of $P_{\rm{T}}$ given by
\begin{align}\label{eq:f4}
\cP^{(\ell)}_6:\quad  \max \limits_{\PT}  \quad & \REbar({\PT}) = \left( {\frac{1}{{\xi \PT + M\Pc + \Ps}}+\frac{\beta}{{\Ptot}}} \right) \ntb
&\qquad\cdot \SEbar( {\PT} ) \ntb
{\mathrm{s.t.}}\quad
& 0 \le \PT \le \Pmax
\end{align}
where $\REbar({\PT})$ is an auxiliary function. Denoting by $P_{\rm{T}}^{*}$ the optimal solution of $\cP^{(\ell)}_6$, we can then obtain that $\REbar({\PT}^*)$ is indeed the optimal objective value of $\cP_4^{(\ell)}$. Based on this fact, we solve the RE optimization problem $\cP^{(\ell)}_4$ via first solving the SE optimization problem $\cP^{(\ell)}_5$ to obtain $\SEbar( {\PT} )$, and then optimizing the parametric problem $\cP^{(\ell)}_6$ to acquire the optimal $P_{\rm{T}}^{*}$. To provide an insight into $\cP^{(\ell)}_6$, we summarize some properties of $\REbar({\PT})$ in the following proposition.
\begin{prop}\label{theorem:properties_RE}
Given a certain overall transmit power $\PT$, $\REbar({\PT})$ is the corresponding maximal system RE in the $\ell$th iteration of the MM method, which has the properties as follows:
\begin{description}
\item[(i)] $\REbar( {\PT} )$ is continuously differentiable and strictly quasi-concave with respect to $\PT$;
\item[(ii)] The derivative of $\REbar( {\PT} )$ over $\PT$ is given by
\begin{align}\label{eq:REbargradient}
&\frac{{\de \REbar \left( {\PT} \right)}}{{\de \PT}} \ntb
& = \frac{{\left( {1 + \beta \frac{{\xi \PT + M{\Pc} + \Ps}}{{\Ptot}}} \right)\frac{{\de \SEbar \left( {\PT} \right)}}{{\de \PT}} - \xi \frac{\EEbar (\PT)}{W}}}
{{\xi \PT + M{\Pc} + {\Ps}}}
\end{align}
where
\begin{align}
\EEbar(\PT) = \frac{W{\SEbar(\PT)}}{{\xi \PT + M{\Pc} + \Ps}}
\end{align}
and the derivative $\frac{{\de \SEbar \left( {\PT} \right)}}{{\de\PT}}$ is given by the optimal Lagrangian multiplier $\mu^*$ related to the power constraint in the SE maximization problem $\cP^{(\ell)}_5$.
\end{description}
\end{prop}
\begin{IEEEproof}
See \appref{app:C}.
\end{IEEEproof}

\propref{theorem:properties_RE} illustrates the strict quasi-concavity and differentiability of $\REbar( {\PT} )$ over $\PT$. Since a unique globally optimal point exists for any strictly quasi-concave problem, Property (i) in \propref{theorem:properties_RE} ensures the existence of a unique global optimum of $\cP^{(\ell)}_6$.
Thus, we can obtain that either $\REbar( {\PT} )$ is non-decreasing in $[0,\Pmax]$, or there exists a point ${P_{\rm{opt}}^{(\ell)}}\in[0,\Pmax]$ that maximizes $\REbar( {\PT} )$ such that $\REbar( {\PT} )$ is monotonically non-decreasing when $P_{\rm{T}} < {P_{\rm{opt}}^{(\ell)}}$, and monotonically non-increasing when $P_{\rm{T}} > {P_{\rm{opt}}^{(\ell)}}$ \cite{Boyd04Convex}. Motivated by the above properties, we decompose $\cP^{(\ell)}_4$ into two-layer nested problems and alternately solve them. Specifically, the decomposition of $\cP^{(\ell)}_4$ can be described as
\begin{description}
\item[(i)] Inner-layer: Solve SE maximization problem $\cP_5^{(\ell)}$ for a given $\PT$, to obtain its maximum $\SEbar\left(\PT\right)$ and the derivative $\frac{{\de \SEbar \left( {\PT} \right)}}{{\de\PT}}$.
\item[(ii)] Outer-layer: Solve RE maximization problem $\cP_6^{(\ell)}$ to obtain the optimal $P_{\rm{opt}}^{(\ell)} \in [0, \Pmax]$ through a derivative-assisted gradient approach according to \propref{theorem:properties_RE}.
\end{description}

Based on \propref{theorem:properties_RE}, it is clear to perform the derivative-assisted gradient approach in the outer-layer optimization. More specifically, given an initial transmit power $\PT(0)$, the optimum of $\cP_6^{(\ell)}$ can be acquired via updating $\PT (t)$ with the derivative of $\REbar\left(\PT\right)$ as follows
\begin{align}\label{eq:PTupdate}
\PT\left(t\right) = \PT\left(t-1\right)+s\times\frac{{\de \REbar \left( {\PT} \right)}}{{\de \PT}}
\end{align}
where $s$ denotes the step length. Therefore, the key rests with the inner-layer subproblem which aims to find $\SEbar\left(\PT\right)$ and the derivative $\frac{{\de \SEbar \left( {\PT} \right)}}{{\de \PT}}$.

\subsection{Water-Filling Scheme}
Note that $\cP^{(\ell)}_5$ is a constrained SE maximization problem with the purpose of finding $\SEbar\left(\PT\right)$ under a given overall transmit power $\PT$. For the solution to $\cP^{(\ell)}_5$, we introduce the following proposition.
\begin{prop}\label{theorem:sumrate_KKT}
The optimal power allocation strategy to $\cP^{(\ell)}_5$, which is denoted by $\Lda^* \triangleq \left\{ {{\Lda}^*_{1}},{{\Lda}^*_{2}},\ldots ,{{\Lda}^*_{K}} \right\}$, is the solution to the following concave optimization problem
\begin{align}\label{eq:Rate_short}
\max \limits_{{\bLambda}} \quad & \sum\nolimits_{k = 1}^K \bigg( {\log \det \left( {\bI_{M} + {{\bf{\Gamma }}_k}{\lambdak}} \right)}  \ntb
&\qquad + \log \det \left( {{{{\bf{\widetilde \Gamma }}}_k} + {{{\bf{\overline K}}}_k \left(\Lda\right)}} \right)
 - \tr { {{{\bf{D }}^{(\ell)}_{k}}{\lambdak}} } \bigg) \ntb
{\mathrm{s.t.}}\quad
& \sum\nolimits_{k=1}^{K}{\tr { \lambdak }} = \PT \ntb
& \lambdak \succeq \bzero,\; \lambdak\; \mathrm{diagonal},\; \forall k\in \K.
\end{align}
The $m$th element of ${\bLambda_{k}^{*}}$ in the solution to \eqref{eq:Rate_short} which is denoted as ${\lambda_{k,m}^{*}}$ satisfies \eqref{eq:Solution_Rate}, shown at the top of this page,
\begin{figure*}
\begin{align}\label{eq:Solution_Rate}
\left\{ \begin{array}{l}
\frac{{\gamma _{k,m}^{*}}}{{1 + \gamma _{k,m}^{*}\lambda _{k,m}^{*}}} + \sum\nolimits_{k' \ne k}^K {\sum\nolimits_{n = 1}^{N_{k'}} {\frac{{{{\widehat r}_{k',m,n}}}}{{\widetilde \gamma _{k',n}^{*} + {\sigma}^2 + \tr {{{\widehat {\bf{R}}}_{k',n}}\bLambda_{\backslash k'}^{*}}}}} }  = d ^{(\ell)}_{k,m} + {\mu ^{*}},\quad{\mu ^{*}} < \upsilon _{k,m}^{*}\\
\lambda _{k,m}^{*} = 0 \quad\quad\quad\quad\quad\quad\quad\quad\quad\quad\quad\quad\quad\quad\quad\quad\quad\;\quad\qquad\qquad\ \ ,\quad {\mu ^{*}} \ge \upsilon _{k,m}^{*}
\end{array} \right.
\end{align}
\hrule
\end{figure*}
where the Lagrange multiplier ${\mu ^{*}}$ satisfies the following KKT conditions
\begin{align}\label{eq:mu_constraint}
{\mu ^{*}}\left( {\tr { {\sum\nolimits_{k=1}^K {\bLambda_k^{^{*}}} } } - \PT} \right) & = 0 \ntb
{\mu ^{*}} & \geq 0
\end{align}
and the auxiliary variable $\upsilon _{k,m}^{*}$ in \eqref{eq:Solution_Rate} is given by
\begin{align}
& \upsilon _{k,m}^{*}  = \gamma _{k,m}^{*}  - d^{(\ell)} _{k,m} \ntb
&\; + \sum\nolimits_{k' \ne k}^K {\sum\nolimits_{n = 1}^{N_{k'}} {\frac{{{{\widehat r}_{k',m,n}}}}{{\widetilde \gamma _{k',n}^{*} + {\sigma}^2 + \sum\limits_{\scriptstyle \ \ (l',m')\hfill\atop
\scriptstyle \in {\cal{S}}(k,m,k')\hfill} {{{\widehat r}_{k',m',n}}\lambda _{l',m'}^{*}} }}} } \ntb
& \quad {{\cal{S}}_{k,m,k'}} = \left\{ {(l',m')|l' \ne k',(l',m') \ne (k,m),} \right.\ntb&\qquad\qquad\quad\left. {l' \in \{ 1,2, \ldots ,K\} ,m' \in \{ 1,2, \ldots ,M\} } \right\}
\end{align}
where $\gamma _{k,m}^{*}$, ${{\widehat r}_{k',m,n}}$, and ${d ^{(\ell)}_{k,m}}$ are the $m$th diagonal elements of ${\bf{\Gamma }}_{k}^{*}$, ${{\widehat {\bf R}}_{k',n}}$ and ${{\bf{D }}^{(\ell)}_{k}}$, respectively, and $\widetilde \gamma _{k',n}^{*}$ is the $n$th diagonal element of ${\bf{\widetilde \Gamma }}_{k'}^{*}$.
\end{prop}
\begin{IEEEproof}
See \appref{app:D}.
\end{IEEEproof}

Notice that the power allocation solution to \eqref{eq:Solution_Rate} resembles the classical water-filling result with the Lagrange multiplier $\mu^*$ in \eqref{eq:mu_constraint} being the water level. In addition, since a sum power constraint is considered, the water levels of all UTs must be equal. Specifically, in the single-UT scenario with $K = 1$, the solution yields a standard water-filling behaviour, thus can be obtained in a closed form, i.e., $\lambda _{k,m}^{*} = {\left[ {{{({d ^{(\ell)}_{k,m}} + {\mu ^{*}})}^{ - 1}} - {{(\gamma _{k,m}^{*})}^{ - 1}}} \right]^ + }$, where the choice of ${\mu ^{*}}$ depends on the constraints in \eqref{eq:mu_constraint}.

Our statistical CSI aided transmission design based on the MM method and deterministic equivalent theory is detailedly presented in \alref{alg:a1}, where $\overline{\eta}_{\mathrm{RE}}^{(\ell)}$ is given by
\begin{align}\label{eq:DE_RE}
\overline{\eta}_{\mathrm{RE}}^{(\ell)} &= \left( {\frac{1}{{\xi \sum\nolimits_{k}{\tr { \lambdak^{(\ell)} } } + M\Pc + \Ps}} + \frac{\beta}{{\Ptot}}} \right)  \ntb
&\qquad\cdot{\sum\nolimits_{k=1}^{K}{ \left( \overline{g}_{k}\left(\Lda^{(\ell)}\right) - f_k \left(\Lda^{(\ell)}\right)\right) }}.
\end{align}
Since it is in general difficult to obtain the solution to \eqref{eq:Solution_Rate} in a closed form for the case with multiple UTs, we also propose a SE maximization iterative water-filling algorithm applied in \alref{alg:a1} to efficiently solve \eqref{eq:Solution_Rate}, which is described in \alref{alg:a2}, where the auxiliary variables $\rho _{k,m,(i)}^{(\ell)}({\overline x_{k,m}})$, ${\rho '}_{k,m,(i)}^{(\ell)}({\overline x_{k,m}})$, and $\mu_{\max}$ are given by
\begin{align}
&\nu _{k,m}^{(\ell)}({\overline x_{k,m}})  =  \frac{{\gamma _{k,m}^{(\ell)}}}{{1 + \gamma _{k,m}^{(\ell)}{\overline x_{k,m}}}} - d^{(\ell)} _{k,m} - \mu \ntb
& +  \sum\nolimits_{k' \ne k}^K  \sum\nolimits_{n = 1}^{N_{k'}} \ntb
&{\frac{{{{\widehat r}_{k',m,n}}}}{{\widetilde \gamma _{k',n}^{(\ell)} + {\sigma}^2 +  {{{\widehat r}_{k',m,n}}{\overline x_{k,m}}} +  \sum\limits_{\scriptstyle \ \ (l',m')\hfill\atop
\scriptstyle \in S(k,m,k')\hfill} {{{\widehat r}_{k',m',n}}x_{l',m'}} }}}  \label{eq:nu_1} \\
&{\nu '}_{k,m}^{(\ell)}({\overline x_{k,m}})  = - \frac{{{{(\gamma _{k,m}^{(\ell)})}^2}}}{{{{(1 + \gamma _{k,m}^{(\ell)}{\overline x_{k,m}})}^2}}}
 -  \sum\nolimits_{k' \ne k}^K \sum\nolimits_{n = 1}^{N_{k'}} \ntb
& {\frac{{\widehat r_{k',m,n}^2}}{{{{(\widetilde \gamma _{k',n}^{(\ell)} + {\sigma}^2 +  {{\widehat r}_{k',m,n}}{\overline x_{k,m}} + \sum\limits_{\scriptstyle \ \ (l',m')\hfill\atop
\scriptstyle \in S(k,m,k')\hfill} {{{\widehat r}_{k',m',n}}x_{l',m'}} )}^2}}}} \label{eq:nu_2} \\
&\mu_{\max}  = \underset{k,m} \max \ \gamma _{k,m}^{(\ell)} +  \sum\nolimits_{k' \ne k}^K {\sum\nolimits_{n = 1}^{N_{k'}} {\frac{{{{\widehat r}_{k',m,n}}}}{{\widetilde \gamma _{k',n}^{(\ell)} + {\sigma}^2}} - d ^{(\ell)}_{k,m}} }\label{eq:mu_max}
\end{align}
respectively.

\begin{algorithm}[h]
\caption{Power Allocation Algorithm for RE Maximization}
\label{alg:a1}
\begin{algorithmic}[1]
\Require Initial power allocation matrices $\{{\Lda}^{(0)}_{k}\} _{k = 1}^K$.
\Ensure Power allocation matrices $\{{\Lda}_{k}\} _{k = 1}^K$.
\State Initialization: threshold ${\varepsilon}_1$, ${\varepsilon}_2$, ${\varepsilon}_3$, iteration index $\ell = 0$.
\State Calculate $\overline{\eta}_{\mathrm{RE}}^{(0)}$ by \eqref{eq:DE_RE}.
\Repeat
\State Initialization: ${\bf{\widetilde \Phi }}_k^{(0)}$, iteration index $u = 0$.
\Repeat
\State Calculate ${\bf{\widetilde \Phi }}_k^{(u+1)}$ and ${\bf{\Phi }}_k^{(u+1)}$ by \eqref{eq:phi_1} and \eqref{eq:phi_2}.
\State Set $u=u+1$.
\Until{$\left|  {\bf{\widetilde \Phi }}_k^{(u)} - {\bf{\widetilde \Phi }}_k^{(u-1)}  \right|\le \varepsilon_1 $}
\State Calculate ${\bf{ \Gamma }}^{(\ell)}_{k}$ and ${\bf{\widetilde \Gamma }}^{(\ell)}_{k}$ by \eqref{eq:gamma_1} and \eqref{eq:gamma_2}, $k = 1,2, \ldots ,K$.
\State Calculate the derivative ${{\bf{D }}^{(\ell)}_{k}}$ by \eqref{eq:derivative_CCCP}, $k = 1,2, \ldots ,K$.
\State Initialization: transmit power $\PT(0)\in[0,\Pmax]$, step length $s$, and iteration index $t=0$.
\Repeat
\State Solve $\cP_5^{(\ell)}$ with $\PT(t)$ to obtain $\SEbar\left(\PT(t)\right)$ and $\frac{{\de \SEbar \left( {\PT(t)} \right)}}{{\de \PT}}$ by \alref{alg:a2}, and set $\{\upnot{\Lda}{temp}_{k}\} _{k = 1}^K$ as $\{ {\Lda}^*_{k,\left(\ell\right)}\} _{k = 1}^K$ which is the output of \alref{alg:a2}.
\State Calculate the derivative $\frac{{\de \REbar \left( {\PT(t)} \right)}}{{\de \PT}}$ by \eqref{eq:REbargradient}.
\State Update $\PT(t+1)$ by \eqref{eq:PTupdate} and set $t=t+1$.
\Until{$\left|  \PT(t) - \PT(t-1)  \right| \le \varepsilon_2$}
\State Calculate $\overline{\eta}_{\mathrm{RE}}^{(\ell+1)}$ by \eqref{eq:DE_RE}.
\State Update $\{ {\Lda}^{(\ell+1)}_{k}\} _{k = 1}^K = \{\upnot{\Lda}{temp}_{k}\} _{k = 1}^K $ and set $\ell=\ell+1$.
\Until{$\left|  \overline{\eta}_{\mathrm{RE}}^{(\ell)} - \overline{\eta}_{\mathrm{RE}}^{(\ell-1)}  \right|\le \varepsilon_3$}
\State Return $\{{\Lda}_{k}\} _{k = 1}^K = \{\Lda_k^{(\ell)}\} _{k = 1}^K$.
\end{algorithmic}
\end{algorithm}

\begin{algorithm}[h]
\caption{Iterative Water-Filling Algorithm for SE Maximization}
\label{alg:a2}
\begin{algorithmic}[1]
\Require Power allocation matrices $\{{\Lda}^{(\ell)}_{k}\} _{k = 1}^K$ and transmit power $\PT(t)$.
\Ensure Power allocation matrices $\{{\Lda}^*_{k}\} _{k = 1}^K$, maximal SE value $\SEbar\left(\PT(t)\right)$, and the derivative $\frac{{\de \SEbar \left( {\PT(t)} \right)}}{{\de \PT}}$.
\State Initialization: diagonal matrices ${{\bf{X}}}_k = {\bLambda_k^{(\ell)}}$ for $k=1,2,\ldots,K $, transmit power $\PT = \PT(t)$, threshold $\varepsilon_4$ and $\varepsilon_5$, Lagrange multiplier ${\mu}^{(u')}_{\min}=0$, and iteration index $u'=0$. Denote $x_{k,m}$ as the $m$th diagonal entry of ${\bf{X}}_k$.
\State Calculate ${\mu}^{(u')}_{\max}$ by \eqref{eq:mu_max} and ${\mu ^{(u')}}= \frac{1}{2} ({\mu^{(u')} _{\max}}+{\mu^{(u')} _{\min}})$.
\Repeat
\For{$k=1$ to $K$}
\For{$m=1$ to $M$}
\State Set ${w_{k,m}}=0$ and $\overline x_{k,m}^{(w_{k,m})}=x_{k,m}$.
\Repeat
\State Calculate $\nu _{k,m}^{(\ell)}(\overline x_{k,m}^{({w_{k,m}})})$ and ${\nu  '}_{k,m}^{(\ell)}(\overline x_{k,m}^{({w_{k,m}})})$ by \eqref{eq:nu_1} and \eqref{eq:nu_2}.
\State Update $\overline x_{k,m}$ as $\overline x_{k,m}^{({w_{k,m}} + 1)} = \overline x_{k,m}^{({w_{k,m}})} - \nu _{k,m}^{(\ell)}(\overline x_{k,m}^{({w_{k,m}})})/{\nu '}_{k,m}^{(\ell)}(\overline x_{k,m}^{({w_{k,m}})})$.
\State Set $w_{k,m}=w_{k,m}+1$.
\Until{$\left| {\overline x_{k,m}^{({w_{k,m}})} - \overline x_{k,m}^{({w_{k,m}} - 1)}} \right| \le {\varepsilon _4}$}
\EndFor
\EndFor
\State Update ${{x}_{k,m}} = {\left[ {\overline x_{k,m}^{({w_{k,m}})}} \right]^ + }$ and calculate ${p_{\rm{tot}}} = \sum\limits_{k = 1}^K {\sum\limits_{m = 1}^M {{x_{k,m}}} } $.
\State Update $ \mu^* = {\mu^{(u')}} $.
\If{${p_{\mathrm{tot}}} < \PT $}
\State Set ${\mu _{\min}^{(u' + 1)}}={\mu _{\min}^{(u')}}$ and ${\mu _{\max}^{(u' + 1)}}={\mu^{(u')}}$.
\Else
\State Set ${\mu _{\min}^{(u' + 1)}}={\mu^{(u')}}$ and ${\mu _{\max}^{(u' + 1)}}={\mu _{\max}^{(u')}}$.
\EndIf
\State Update ${\mu ^{(u' + 1)}}= \frac{1}{2} ({\mu _{\max}^{(u' + 1)}}+{\mu _{\min}^{(u' + 1)}})$ and set $u'=u'+1$.
\Until{$\left| {\PT - {p_{\mathrm{tot}}}} \right| \le {\varepsilon _5}$}
\State Return $\{{\Lda}^*_{k}\} _{k = 1}^K = \{\mathbf{X}_{k}\} _{k = 1}^K$, $\SEbar\left(\PT(t)\right) = \sum\nolimits_k { \left( \Rbar_{k}\left(\Lda^*\right) - \triangle f_{k,\ub}^{(\ell)} \left(\Lda^*\right) \right) }$, and $\frac{{\de \SEbar \left( {\PT(t)} \right)}}{{\de \PT}} = \mu^*$.
\end{algorithmic}
\end{algorithm}

\begin{remark}
The generalized water-filling \alref{alg:a2} can be considered as an extension of the classical water-filling algorithm to our considered multi-UT case, in which the summation of fractional functions poses great difficulty in solving \eqref{eq:Solution_Rate} to obtain accurate solutions. To overcome this, we compute approximate roots of \eqref{eq:Solution_Rate} using the iterative Newton-Raphson method \cite{cormen2009introduction} in Step 9. In addition, the bisection approach is exploited to search for  ${\mu ^{*}}$ under the constraints in \eqref{eq:mu_constraint}. For the single-UT case, substituting the explicit solution to \eqref{eq:Solution_Rate} for the approximate solution obtained by Newton-Raphson method, \alref{alg:a2} reduces to the standard water-filling algorithm.
\end{remark}

\begin{remark}
Iterative water-filling approaches are proposed for the SE and EE maximization problems with instantaneous CSIT in \cite{Sum2005Jindal} and \cite{xu2013energy}, respectively, whereas our proposed one applies for the RE optimization problem with statistical CSIT.
\end{remark}

\subsection{Convergence and Complexity Analysis}
For the convergence of the proposed low-complexity algorithms, we start with the convergence of \alref{alg:a2} owing to the use of the SE maximization iterative water-filling procedure in \alref{alg:a1}. For the inner-layer problem of solving $\cP_4^{(\ell)}$, since $\cP_5^{(\ell)}$ is a concave problem, the SE maximization iterative water-filling can achieve the global maximum through solving the KKT optimality conditions \cite{Boyd04Convex}. Thus, the SE maximization iterative water-filling converges to the global optimum for $\cP_5^{(\ell)}$. For the outer-layer problem of solving $\cP_4^{(\ell)}$, since $\cP_6^{(\ell)}$ is a strictly quasi-concave problem, $\REbar( {\PT} )$ either monotonically increases in $[0, \Pmax]$ or first increases and then decreases with $\PT$. Therefore, the proposed derivative-assisted gradient approach will either end with $\Pmax$ when $\REbar( {\PT} )$ is strictly increasing in $[0, \Pmax]$ or converge to the global optimum $\Popt \in [0, \Pmax]$ for $\cP_6^{(\ell)}$. Moreover, based on the properties of the MM method \cite{sun2017majorization}, the proposed low-complexity power allocation \alref{alg:a1} is convergent. Besides, the optimization of $\Lda$ monotonically increases the objective function of the original problem $\cP_1$ at each iteration, and so does the alternating optimization method. Thus, the overall method that alternatively optimizes $\lambdak(\forall k)$ and $\mathbf{\Psi}_k(\forall k)$ is guaranteed to converge.

Then, we discuss the complexity of the proposed algorithms. For each iteration of the MM method in \alref{alg:a1}, the number of iterations involved in calculating ${\bf{\widetilde \Phi }}_k^{(u+1)}$ and ${\bf{ \Phi }}_k^{(u+1)}$ depends on the predefined threshold, and is usually very small in the numerical experiments.
Then, the major complexity of the MM method is composed of the complexity of the two-layer scheme to solve problem $\cP_4^{(\ell)}$. Since the derivative-assisted gradient approach applied in the outer layer will converge very fast \cite{Boyd04Convex}, the complexity depends mainly on \alref{alg:a2} for solving the inner-layer problem. For the complexity of \alref{alg:a2}, the number of iterations required in the convergence of Newton-Raphson method to solve \eqref{eq:Solution_Rate} is dominant since the bisection method has a relatively fast convergence rate \cite{Boyd04Convex}. Thus, the total computational complexity of \alref{alg:a1} is approximately $\cO (L_{\mathrm{M}}L_{\mathrm{G}}L_{\mathrm{N}}KM + L_{\mathrm{M}}KM)$, where $L_{\mathrm{M}}$, $L_{\mathrm{G}}$, and $L_{\mathrm{N}}$ are the numbers of iterations required for the MM method, the derivative-assisted gradient approach, and Newton-Raphson method, respectively. Note that the values of $L_{\mathrm{M}}$, $L_{\mathrm{G}}$, and $L_{\mathrm{N}}$ will depend on the preset thresholds.

\section{Numerical Results}\label{sec:numerical_results}

\begin{figure}
\centering
\includegraphics[width=\figdoucolwid]{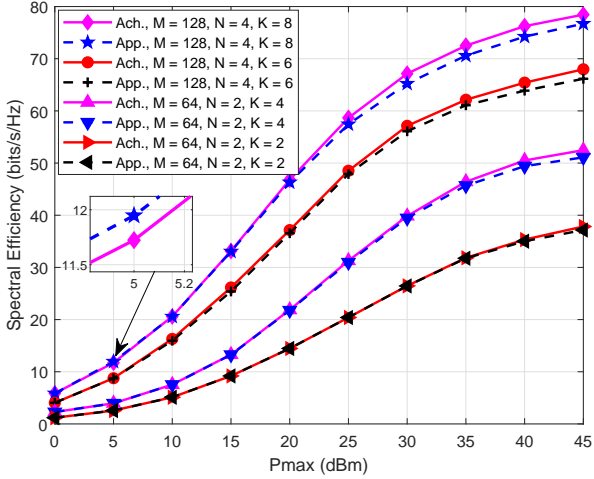}
\caption{Comparison between the achievable ergodic rate expression in \eqref{eq:ergodic_rate1} and the approximated rate expression in \eqref{eq:ergodic_rate} under different setup parameters.}
\label{fig:rateapp}
\end{figure}

\begin{figure}
\centering
\includegraphics[width=\figdoucolwid]{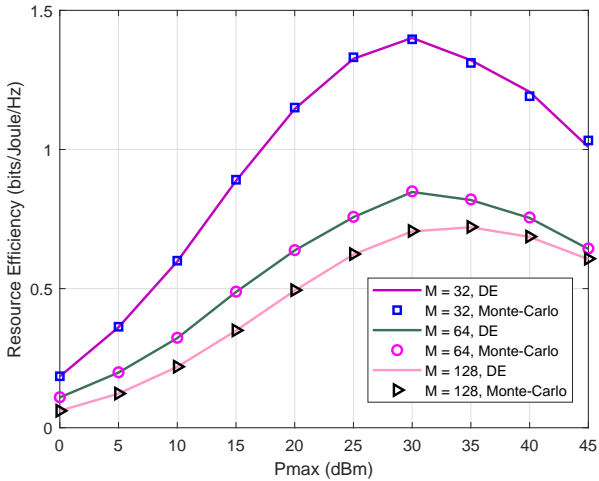}
\caption{The RE performance versus the power budget $\Pmax$ for different numbers of BS antennas $M$ ($\beta = 0.5$).}
\label{fig:M_different}
\end{figure}

Numerical analysis is presented to evaluate the performance of the proposed statistical CSI aided RE optimization framework for massive MIMO downlink transmission. The QuaDRiGa channel model \cite{jaeckel2014quadriga} with a suburban macro cell scenario is adopted throughout the simulations. A total of $K=8$ UTs are randomly distributed in the cell sector. The pathloss is set as $-120$ dB for all UTs \cite{shen2018fractional}. The antenna array topology ULA is adopted for the BS and each UT $k$, with the numbers of antennas being $M=128$ and $N_k = 4(\forall k)$, respectively, and the spacing between antennas is a half wavelength. The bandwidth is set as $W = 10$ MHz. The amplifier inefficiency factor is set as $\xi = 5$, the hardware dissipated power per antenna and the static power consumption are set to $\Pc = 30$ dBm and $\Ps = 40$ dBm, respectively. The background noise variance is set as ${\sigma}^2 = -105$ dBm \cite{he2013coordinated}.

\figref{fig:rateapp} compares the achievable ergodic rate expression in \eqref{eq:ergodic_rate1} with the approximated rate expression in \eqref{eq:ergodic_rate} under different system setup parameters. We can observe that the adopted rate expression is a good approximation with different numbers of BS antennas, UT antennas, UTs, and power budgets.

In \figref{fig:M_different}, we compare the proposed algorithm with Monte-Carlo method where channel sample averaging is utilized to approximate the expectation operation. It can be observed that the DE results are almost identical to those obtained from Monte-Carlo method in all the considered scenarios, which demonstrates the effectiveness of the proposed DE-based optimization framework.

\begin{figure}
\centering
\includegraphics[width=0.45\textwidth]{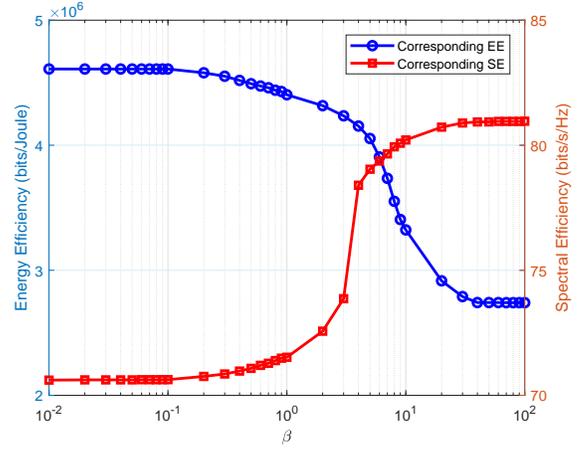}
\caption{Influence of the weighting factor $\beta$ on the corresponding system EE and SE ($\Pmax=45$ dBm, $M=128$).}
\label{fig:corresponding}
\end{figure}

\begin{figure}[!t]
\centering
\includegraphics[width=\figdoucolwid]{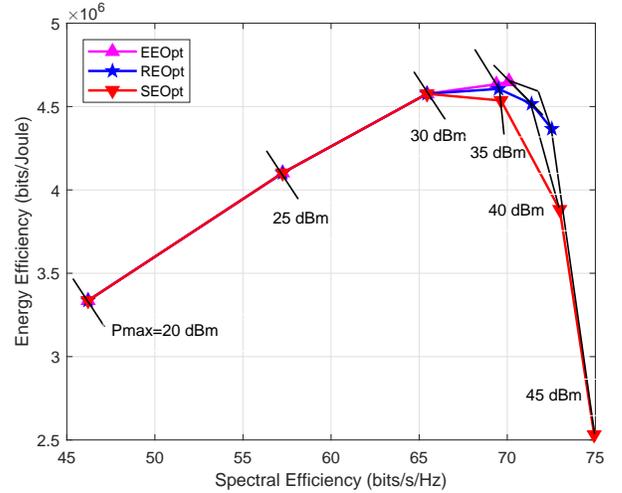}
\caption{EE-SE tradeoff under different transmit power budgets ($\beta\frac{ W}{\Ptot} = 0.5$).}
\label{fig:tradeoff}
\end{figure}

\figref{fig:corresponding} demonstrates the influence of the weighting factor $\beta$ through illustrating the corresponding system EE and SE versus different values of $\beta$.
The results indicate that increasing $\beta$ results in an improved system SE but a reduced system EE. This is due to the reason that a larger $\beta$ gives a higher priority to SE and thus allocating more power to maximize SE. Moreover, when $\beta \to 0$, the RE maximization approach reduces to the EE maximization approach, and when $\beta \to \infty$, it tends to maximize the system SE. Generally, \figref{fig:corresponding} reveals the ability of the proposed RE optimization (REOpt) approach for balancing the tradeoff between EE and SE via selecting a suitable weighting factor $\beta$.

\begin{figure}[!t]
\centering
\subfloat[]{\centering\includegraphics[width=0.5\textwidth]{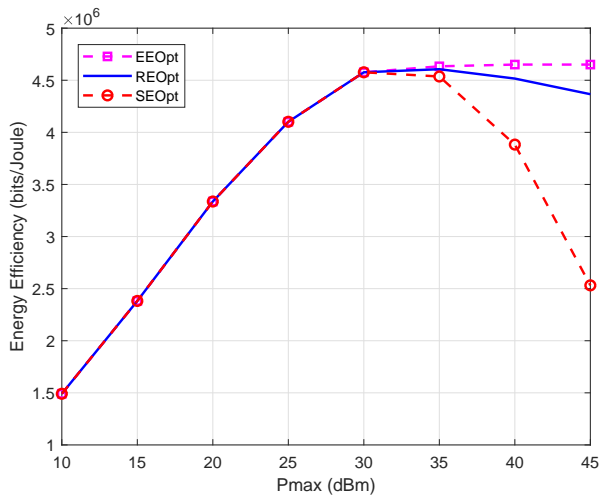}
\label{fig:EE_performance}}
\\
\subfloat[]{\centering\includegraphics[width=0.5\textwidth]{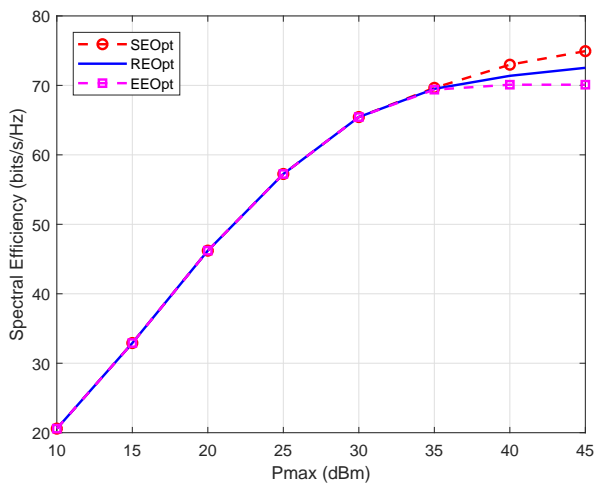}
\label{fig:Rate_performance}}
\caption{Comparison between the EE and SE performance versus $\Pmax$ ($\beta\frac{W}{\Ptot} = 0.5$). (a) EE performance; (b) SE performance.}
\label{fig:Comparison}
\end{figure}

\figref{fig:tradeoff} illustrates the EE-SE tradeoff under different transmit power budgets obtained by the proposed approach. For comparison, we also plot the corresponding performance of the EE optimization (EEOpt) and SE optimization (SEOpt) approaches. The results exhibit that the REOpt approach can balance the EE and SE while the conventional EEOpt/SEOpt one takes only a single criterion into account.
To further show the effectiveness of the proposed REOpt approach, Figs. \ref{fig:Comparison}(a) and \ref{fig:Comparison}(b) evaluate the corresponding EE and SE performance of the three approaches versus the transmit power budget $\Pmax$. In the low transmit power budget regime, we can observe that the three considered approaches exhibit almost identical performance, and both EE and SE can be maximized when $\Pmax \le 30$ dBm, which indicates that transmission with full power budget can achieve a near-optimal balance in the low transmit power budget regime. In the large transmit power budget regime, \alref{alg:a1} achieves neither optimal EE nor optimal SE performance, but strike the balance between the EE and SE, which is in accord with the purpose of our RE maximization design to balance the tradeoff between EE and SE.

\begin{figure}
\centering
\includegraphics[width=\figdoucolwid]{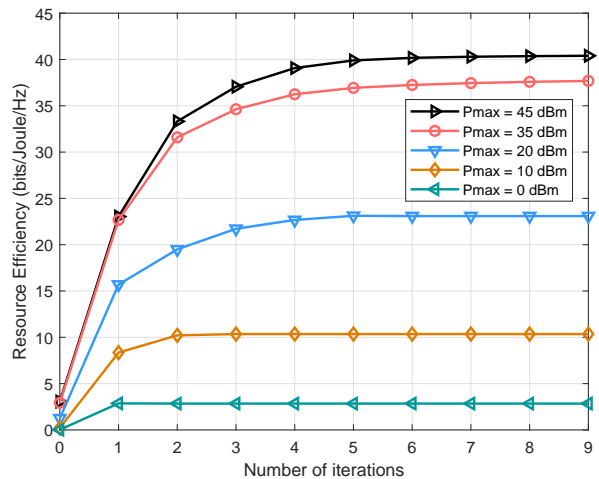}
\caption{Convergence behavior of \alref{alg:a1} versus the number of iterations for different values of transmit power budget $\dnnot{P}{max}$.}
\label{fig:Convergence}
\end{figure}

\begin{figure}
\includegraphics[width=\figdoucolwid]{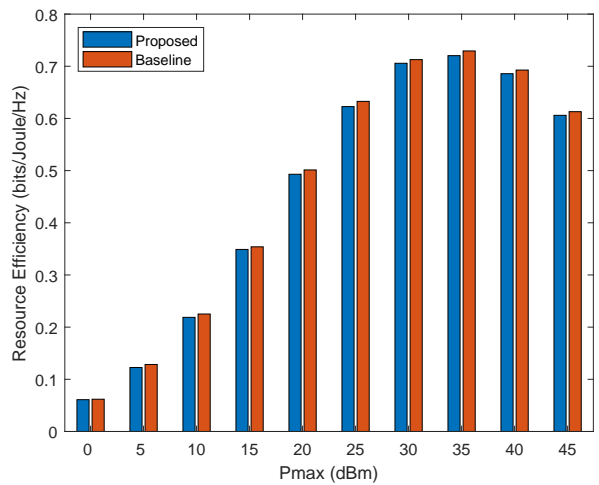}
\centering
\caption{RE performance comparison between the proposed approach and the baseline ($\beta = 0.5$).}
\label{fig:baseline}
\end{figure}

\figref{fig:Convergence} presents the convergence behavior of the iterative \alref{alg:a1} versus the numbers of iterations of the MM method under different transmit power budgets. The results indicate that the proposed \alref{alg:a1} generates a non-decreasing RE value sequence and converges fast in typical transmit power budget regions. In particular, the system RE converges after only one or two iterations in cases of low transmit power budgets. We can also observe that the convergence rate of \alref{alg:a1} becomes slightly slower when $\Pmax$ increases because more iterations are required to find the convergence point in a larger constraint set for higher $\Pmax$.

In \figref{fig:baseline}, we compare the proposed approach with a baseline one, which is obtained by utilizing the proposed algorithms over different initializations and then selecting the best solution. We can observe that RE performance gap between the proposed approach and the baseline can be almost neglected.

\section{Conclusion}\label{sec:conclusion}
We have investigated single-cell massive MIMO downlink precoding under the RE maximization criterion with statistical CSIT. We first showed the solution of the optimal transmit signal direction in a closed form, which indicated that massive MIMO downlink transmission for RE maximization should be performed in the beam domain. Based on this insight, we reduced the complex transmit strategy design into a real-valued power allocation problem in the beam domain. Exploiting the MM method, a suboptimal sequential algorithm was further proposed to solve such a power allocation problem, together with the reduction of computational complexity using random matrix theory. Moreover, we proposed a two-layer scheme to solve each subproblem in the MM method relying on the derivative-assisted gradient approach and generalized iterative water-filling approach. We demonstrated by numerical results the performance gain of the proposed RE maximization approach over the conventional approaches, especially in the high transmit power budget regime.

\appendices

\section{Proof of \propref{theorem:beam_domain_optimal}}\label{app:A}

Define $\overline{\bQ}_k = \bV^H \bQ_k \bV(\forall k)$ for notational brevity. Following a similar proof procedure as that in \cite{tulino2006capacity,You2020energy}, we can obtain that $\overline{\bQ}_k$ should be diagonal for all $k$ to maximize $\SE$ in \eqref{eq:SE_definition} and $\EE$ in \eqref{eq:EE_definition}. Besides, the off-diagonal elements of $\overline{\bQ}_k(\forall k)$ do not affect the value of $\sum\nolimits_{k} \tr{{\bf{Q}}_{k}}$ and thus do not affect the power consumption term $P_{\mathrm{sum}}$ in \eqref{eq:power_consumption_model}. As a result, we can conclude that to maximize the objective of problem $\cP$ under the given constraints in \eqref{eq:problem1}, ${\bf{V}}^H {\bf{Q}}_{k} {\bf{V}}(\forall k)$ should be diagonal, i.e., $\bQ_k^{\mathrm{opt}} = \bV \bLambda_k \bV^H(\forall k)$.

\section{Proof of \propref{prop:mm_convergence}}\label{app:B}
Consider a maximization problem $\mathcal{F} = \underset{\mathbf{x} \in \mathcal{X}} \max \ { g(\mathbf{x}) } $ with the feasible set $\mathcal{X}$ being convex and closed. The MM method aims to find a series of approximation subproblems of $\mathcal{F}$, which can be relatively simpler to handle. More formally, denote by ${\mathcal{F}}^{(\ell)} = \underset{\mathbf{x} \in \mathcal{X}} \max \ { g^{(\ell)} (\mathbf{x}) }$ the $\ell$th subproblem in the MM method with ${\mathbf{x}^{(\ell)}}$ being its maximizer. Then, the following properties are satisfied for all $\ell$
\begin{description}
\item[$\mathbf{P1}$:] ${g^{(\ell)}}(\mathbf{x}) \le g(\mathbf{x}),\ \forall \mathbf{x}$
\item[$\mathbf{P2}$:] ${g^{(\ell)}}({\mathbf{x}^{(\ell - 1)}}) = g({\mathbf{x}^{(\ell - 1)}})$
\item[$\mathbf{P3}$:] $\nabla {g^{(\ell)}}({\mathbf{x}^{(\ell - 1)}}) = \nabla g({\mathbf{x}^{(\ell - 1)}})$.
\end{description}
The sequence $\left\{ g\left({\mathbf{x}^{(\ell)}}\right)\right\}_{\ell=0}^{\infty}$ is the objective value of the original problem $\mathcal{F}$ corresponding to $\left\{{\mathbf{x}^{(\ell)}}\right\}_{\ell=0}^{\infty}$, which is the solution to the subproblem sequence $\left\{{\mathcal{F}}^{(\ell)}\right\}_{\ell=0}^{\infty}$.
If the properties $\mathbf{P1}$, $\mathbf{P2}$, and $\mathbf{P3}$ described above are satisfied, we can obtain that $\left\{g\left({\mathbf{x}^{(\ell)}}\right)\right\}_{\ell=0}^{\infty}$ converges and every limit point of $\left\{{\mathbf{x}^{(\ell)}}\right\}_{\ell=0}^{\infty}$ is a KKT point of $\mathcal{F}$ \cite{razaviyayn2013unified}. In addition, the resulting point ${\mathbf{x}}^*$ satisfies the KKT conditions of $\mathcal{F}$ \cite{marks1978general}.

Now, we consider our RE maximization problem $\cP_3$. Since $f_k$ is concave, the inequality $f_k\left(\Lda\right) \le \triangle f_{k,\mathrm{ub}}^{\left(\ell\right)}\left(\Lda\right)$ holds for $\forall \Lda$. We can then obtain
\begin{align}
&  \sum_{k=1}^{K} \left( \Rbar_k \left( \Lda \right) - \triangle f_{k,\mathrm{ub}}^{\left(\ell\right)}\left(\Lda\right) \right)
\leq
\sum_{k=1}^{K} \left( \Rbar_k \left( \Lda \right) -  f_{k}\left(\Lda\right) \right).
\end{align}
Thus, Property $\mathbf{P1}$ can be shown to be satisfied in terms of $\cP_4^{(\ell)}$. In addition, it is not difficult to show that Properties $\mathbf{P2}$ and $\mathbf{P3}$ are both satisfied.
Moreover, utilizing the concavity of $f_k \left(\Lda\right)$ over $\Lda$, we obtain the following inequality for all $\ell$
\begin{subequations}
\begin{align}\label{eq:gea}
&  \sum_{k=1}^{K} \left( \Rbar_k \left( {\Lda}^{(\ell+1)} \right) - f_k \left( {\Lda}^{(\ell+1)} \right) \right) \\ \label{eq:ge1}
\ge \ & \sum\nolimits_{k=1}^{K} \left( \Rbar_k \left( {\Lda}^{(\ell+1)} \right) - f_k \left( {\Lda}^{(\ell)} \right) \right.\ntb&\left.\qquad\qquad- {\tr { {{{{\bf{D }}_{k}^{(\ell)}}} \left( {{\bLambda^{(\ell+1)}_k} - {\bLambda_{k}^{(\ell)}}} \right)} }} \right)\\ \label{eq:ge2}
\ge \ & \sum\nolimits_{k=1}^{K} \left( \Rbar_k \left( {\Lda}^{(\ell)} \right) - f_k \left( {\Lda}^{(\ell)} \right) \right.\ntb&\left. \qquad\qquad-{\tr { {{{{\bf{D }}_{k}^{(\ell)}}} \left( {{\bLambda^{(\ell)}_k} - {\bLambda_{k}^{(\ell)}}} \right)} }} \right)\\
= \ & \sum\nolimits_{k=1}^{K} \left(\Rbar_k \left( {\Lda}^{(\ell)} \right) - f_k \left( {\Lda}^{(\ell)} \right) \right)
\end{align}
\end{subequations}
where ${\Lda}^{(\ell)}$ is the power allocation result in the $\ell$th iteration. The inequality in \eqref{eq:ge1} follows from the concavity of $ f_k \left(\Lda\right)$. The inequality in \eqref{eq:ge2} is due to the fact that ${\Lda}^{(\ell+1)}$ is the optimum of the maximization problem $\cP_4^{(\ell)}$. Then, according to \eqref{eq:gea}, we can obtain that the objective value sequence of $\left\{\cP_4^{(\ell)}\right\}_{\ell=0}^{ \infty }$ is non-decreasing. Consequently, the conclusions in \propref{prop:mm_convergence} hold.

\section{Proof of \propref{theorem:properties_RE}}\label{app:C}
Since $\REbar\left(\PT\right)$ can be written as $\frac{W\SEbar( {\PT} )}{{\xi \PT + M\Pc + \Ps}}+\alpha\SEbar\left(\PT\right)$, where $\alpha=\beta/\Ptot$, we start from showing that $\SEbar\left(\PT\right)$ is continuously differentiable and concave over $\PT$. Note that the first term of the objective function in $\cP_5^{(\ell)}$, i.e., the DE expression $\Rbar_{k} \left( \Lda \right)$, is concave over $\Lda$ \cite{Dumont2010On,Dupuy2011On}, and the second term $f_k(\bLambda)$ is linearized around the solution of the present iteration. Consequently, the objective function in $\cP_5^{(\ell)}$ is concave over $\Lda$. Utilizing the brief notations $q\left(\Lda\right) = \sum\nolimits_k{ \left( \Rbar_{k} \left(\Lda\right) - \Ta \right) }$ and $p\left(\Lda\right) = \sum\nolimits_k {\tr { \lambdak }}$, we reformulate the SE maximization problem $\cP_5^{(\ell)}$ as follows
\begin{align}\label{eq:SEconcave}
\SEbar \left( {\PT} \right) =  \mathop {\max }\limits_{\Lda} \quad &  { q\left(\Lda\right) } \ntb
{\mathrm{s.t.}}\quad
& p\left(\Lda\right) \le \PT.
\end{align}
Note that relaxation of the power constraint $p\left(\Lda\right) \le \PT $ only increases $\SEbar(\PT)$ and meanwhile $\SEbar(\PT)$ increases as $\PT$ increases.

Following a similar approach as that in \cite[Proposition 1]{hellings2013energy}, we show the concavity of $\SEbar(\PT)$ over $\PT$ by performing a sensitivity analysis \cite[Section 5.6.2]{Boyd04Convex} as follows
\begin{subequations}
\begin{align}
\SEbar\left( {\PT} \right)  &= \mathop {\min }\limits_{\mu  \ge 0} \mathop {\max}\limits_{\Lda \succeq \bzero} \quad q \left( {\Lda} \right) - \mu \left( {p\left( {\Lda} \right) - \PT} \right) \label{eq:1} \\
  &\le \quad \quad \mathop {\max}\limits_{ \Lda \succeq \bzero} \quad q \left( {\Lda} \right) - \widetilde{\mu} \left( {p\left( {\Lda} \right) - \PT} \right) \label{eq:2} \\
  & = \quad \quad \mathop {\max}\limits_{\Lda \succeq \bzero}  \quad q \left( {\Lda} \right) - \widetilde{\mu} \left( {p\left( {\Lda} \right) - \PTt} \right) \ntb&\qquad\qquad\qquad + \widetilde{\mu} \left( \PT - \PTt \right) \label{eq:3} \\
  & = \quad  \SEbar\left( {\PTt} \right) + \widetilde{\mu} \left( \PT - \PTt \right) \label{eq:4}
\end{align}
\end{subequations}
where $\widetilde{\mu}$ and $\PTt$ satisfy
\begin{align}
\widetilde{\mu} &\geq 0 \ntb
\widetilde{\mu} \left( { p\left(\Lda\right) - \PTt } \right) &= 0.
\end{align}
In \eqref{eq:1}, since \eqref{eq:SEconcave} is concave, the equality can be obtained through the strong duality \cite[Section 5.2.3]{Boyd04Convex}. In \eqref{eq:2}, owing to the minimization of $\mu$ in \eqref{eq:1}, the inequality holds for $\forall \mu  \ge 0$. In \eqref{eq:4}, the equality follows from the constraint $\widetilde{\mu} \left( { p\left(\Lda\right) - \PTt } \right) = 0$. Note that \eqref{eq:4} gives an upper bound on the concave function in \eqref{eq:SEconcave} in terms of the subgradient $\widetilde{\mu}$ at point $\PTt$. Moreover, \eqref{eq:4} implies that there exists a subgradient in each point $\PTt$, which indicates that $\SEbar\left( {\PT} \right)$ is concave over $\PT$ \cite[Section 6.5.5]{Boyd04Convex}.

Then, we prove the differentiability of $\SEbar\left( {\PT} \right)$ via illustrating that each subgradient $\widetilde{\mu}$ in the given point $\PTt$ is unique.
The lagrangian function of $\cP_5^{(\ell)}$ is defined as
\begin{align}\label{eq:lag1}
{\cal L}  &= { \sum\nolimits_{k}{\left(    \Rbar_{k}\left(\Lda\right) - \Ta  \right)} } \ntb
&+ \sum\nolimits_{k}{\tr {{{\bf{\Psi }}_k}{\bLambda_k}}} - \mu\left( \sum\nolimits_{k}{\tr {{\bLambda_k}}}-\PT \right)
\end{align}
where the Lagrange multipliers ${{\bf{\Psi }}_k}\succeq \bzero (\forall k)$ depend on the problem constraints. The gradient of ${\Rbar}_{k} \left(\Lda\right)$ over $\bLambda_{k}$ can be derived from \eqref{eq:DE} as
\begin{align}\label{eq:g_Rk}
\frac{\partial }{{\partial {\bLambda_k}}}\Rbar_k (\bLambda) & = {\left( {\bI_{M} + {{\bf{\Gamma }}_k}{\bLambda_k}} \right)^{ - 1}}{{\bf{\Gamma }}_k}\\
 & + \sum\limits_{m,n} {\frac{{\partial \Rbar_k (\bLambda)}}{{\partial {{[{{\widetilde \eta }_k}({\bf{\Phi }}_k^{ - 1}{\bLambda_k})]}_{m,n}}}}\frac{{\partial {{[{{\widetilde \eta }_k}({\bf{\Phi }}_k^{ - 1}{\bLambda_k})]}_{m,n}}}}{{\partial {\bLambda_k}}}} \ntb
 & + \sum\limits_{m,n} {\frac{{\partial \Rbar_k (\bLambda)}}{{\partial {{[{\eta _k}({\bf{\widetilde \Phi }}_k^{ - 1}{\overline{\bf{K}}}_k^{ - 1})]}_{m,n}}}}\frac{{\partial {{[{\eta _k}({\bf{\widetilde \Phi }}_k^{ - 1}{\bf{\overline K}}_k^{ - 1})]}_{m,n}}}}{{\partial {\bLambda_k}}}}.
\end{align}
Following a approach similar to that of proving Theorem 4 in \cite{Lu16Free}, we have
\begin{align}\label{eq:der2}
{\frac{{\partial \Rbar_k (\bLambda)}}{{\partial {{[{{\widetilde \eta }_k}({\bf{\Phi }}_k^{ - 1}{\bLambda_k})]}_{m,n}}}}} = 0 \\
{\frac{{\partial \Rbar_k (\bLambda)}}{{\partial {{[{\eta _k}({\bf{\widetilde \Phi }}_k^{ - 1}{\bf{\overline K}}_k^{ - 1})]}_{m,n}}}}}=0
\end{align}
which further leads to
\begin{align}\label{eq:g_k}
\frac{\partial }{{\partial {\bLambda_k}}}\Rbar_{k} \left( \Lda \right)= {\left( {\bI_{M} + {\bf{\Gamma }}_k{\bLambda_k}} \right)^{ - 1}}{\bf{\Gamma }}_k.
\end{align}
In addition, the gradient of $\Rbar_{k} \left(\Lda\right)$ over $\Lda_{k'}(\forall k' \ne k)$ is derived as
\begin{align}\label{eq:g_l}
\frac{\partial }{{\partial {\bLambda_{k'}}}}\Rbar_{k} \left(\Lda\right) = \sum\nolimits_{n = 1}^{N_k} {\frac{{{{\widehat {\bf{R}}}_{k,n}}}}{{\widetilde \gamma _{k,n} + {\sigma}^2 +  \tr {{\bLambda_{\backslash k}}{{\widehat {\bf{R}}}_{k,n}}}}}}
\end{align}
where ${\widetilde \gamma _{k,n}}$ denotes the $n$th diagonal element of ${\bf{\widetilde \Gamma }}_k$. Then, from \eqref{eq:g_k} and \eqref{eq:g_l}, we have
\begin{align}
&\frac{\partial }{{\partial {\bLambda_a}}}\sum\nolimits_{k = 1}^K {\Rbar_k \left(\Lda\right)}  = {\left( {\bI_{M} + {\bf{\Gamma }}_a{\bLambda_a}} \right)^{ - 1}}{\bf{\Gamma }}_a \ntb
&\qquad\quad+ \sum\nolimits_{k \ne a}^K {\sum\nolimits_{n = 1}^{N_k} {\frac{{{{\widehat {\bf{R}}}_{k,n}}}}{{\widetilde \gamma _{k,n} + {\sigma}^2 + \tr {{\bLambda_{\backslash k}}{{\widehat {\bf{R}}}_{k,n}}}}}} }.
\end{align}
Owing to the concavity of $\Rbar_{k} \left(\Lda\right) $ over $\Lda$, the KKT conditions of $\cP_5^{(\ell)}$ are
\begin{align}\label{eq:KKT_1}
 \frac{{\partial {{\cal L}}}}{{\partial \lambdak^* }} & = \bzero, \quad k = 1, \ldots ,K \\ \label{eq:KKT_2}
 \tr {{\bf{\Psi }}_k^* \lambdak^* } & = 0, \quad {\bf{\Psi }}_k^* \succeq \bzero,\quad \lambdak^* \succeq \bzero \\ \label{eq:KKT_3}
 {\mu^*} \left( \sum\nolimits_{k}{\tr {\lambdak^*}}-\PT \right) & =0, \quad {\mu^*}\ge 0.
\end{align}

From \eqref{eq:lag1} and \eqref{eq:der2}, we reformulate the first KKT condition in \eqref{eq:KKT_1} as
\begin{align}\label{eq:solution_matrix}
\frac{{\partial {{\cal L}}}}{{\partial \lambdak^* }}  = \bB_k
 - \bD^{(\ell)}_k +  {\bf{\Psi }}_k^* - \mu^* \bI_{M} = \bf{0}
\end{align}
where
\begin{align}\label{eq:def_bk}
 \bB_k &\triangleq
 {\left( {\bI_{M} + {\bf{\Gamma }}_k^* \lambdak^* } \right)^{ - 1}}{\bf{\Gamma }}_k^* \ntb
 &\quad+  \sum\limits_{k' \ne k}^K {\sum\limits_{n = 1}^{N_{k'}} {\frac{{{{\widehat {\bf R}}_{k',n}}}}{{\widetilde \gamma^* _{k',n} + {\sigma}^2 + \tr {\bLambda^*_{\backslash k'}{{\widehat \bR }_{k',n}}}}}} }.
\end{align}
From \eqref{eq:KKT_2} and \eqref{eq:solution_matrix}, we obtain that
\begin{align}
&\bB_k - \bD^{(\ell)}_k + {\bf{\Psi }}_k^*- \mu^* \bI_M = \bzero\\
&{\bf{\Psi }}_k^* \lambdak^* = \bzero, \quad \forall k
\end{align}
where $\bB_k$ and $\bD^{(\ell)}_k$ defined in \eqref{eq:def_bk} and \eqref{eq:derivative_CCCP}, respectively, are both non-negative definite diagonal matrices. Besides, the Lagrange multiplier matrix ${\bf{\Psi }}_k^*$ and the power allocation matrix $\lambdak^*$ are also both non-negative definite and diagonal. It is infeasible to change $\mu^*$ without changing at least one diagonal element of $\bB_k$, in other words, changing at least one $\lambdak^*$. As a result, there exists a unique multiplier $\mu^*$ satisfying the KKT conditions for the given optimizer $\left\{\lambdak^*\right\}_{k=1}^K$. Note that the objective function of $\cP_5^{(\ell)}$ in \eqref{eq:ratemax} is given by $\sum_{k=1}^{K} { \left( \overline{g}_{k}\left(\bLambda\right) - \triangle f_{k,\mathrm{ub}}^{(\ell)} \left(\bLambda\right) \right) }$. As the DE expression $\overline{g}_{k}\left(\bLambda\right)$ is strictly concave with respect to $\bLambda_k (\forall k)$ \cite{Lu16Free} and the first-order Taylor expansion $\triangle f_{k,\mathrm{ub}}^{(\ell)} \left(\bLambda\right)$ is an affine function, we can obtain that the objective function of problem $\cP_5^{(\ell)}$ is strictly concave with respect to $\bLambda_k (\forall k)$. Therefore, the optimizer $\left\{\lambdak^*\right\}_{k=1}^K$ for the given point $\PT$ is also unique, which further indicates that the optimal $\mu^*$ is unique. Consequently, the subgradient $\widetilde{\mu}$ in the given point $\PTt$ in \eqref{eq:4} is unique. Thus, $\widetilde{\mu}$ is a gradient, which proves that $\SEbar\left(\PT\right)$ is continuously differentiable over $\PT$.

Since $\SEbar \left( {\PT} \right)$ is concave over $\PT$ and $\xi \PT + M\Pc + \Ps$ is linear over $\PT$, $\frac{W\SEbar( {\PT} )}{{\xi \PT + M\Pc + \Ps}}$ is strictly quasi-concave and continuously differentiable over $\PT$. In addition, $\alpha \SEbar(\PT)$ is concave over $\PT$. Thus, $\REbar( {\PT} )$ is strictly quasi-concave and continuously differentiable over $\PT$. This concludes the proof of Property (i) in \propref{theorem:properties_RE}.

Next, we analyze the derivative of $\REbar( {\PT} )$ with respect to $\PT$. Using the chain rule, we first derive the derivative of ${\EEbar\left( {\PT} \right)}$ with respect to $\PT$ as follows
\begin{align}
\frac{{\de{\EEbar\left( {\PT} \right)}}}{{\de\PT}}
&= \frac{{\de{\frac{W{\SEbar(\PT)}}{{\xi \PT + M{\Pc} + \Ps}}}}}{{\de\PT}}  \ntb
&=  W\frac{{\frac{{\de\SEbar\left( {\PT} \right)}}{{\de\PT}}\left(\xi \PT + M\Pc + \Ps\right) - \xi \SEbar\left( {\PT} \right)}}{{\left(\xi \PT + M\Pc + \Ps\right)^2}} \ntb
& = \frac{{W\frac{{\de\SEbar\left( {\PT} \right)}}{{\de\PT}} - \xi \EEbar\left( {\PT} \right)}}{{\xi \PT + M\Pc + \Ps}}.
\end{align}
Then, we have
\begin{align}
\frac{{\de \REbar \left( {\PT} \right)}}{{\de \PT}}& = \frac{{\left( {1 + \beta \frac{{\xi \PT + M{\Pc} + \Ps}}{{\Ptot}}} \right)\frac{{\de \SEbar \left( {\PT} \right)}}{{\de \PT}} - \xi \frac{\EEbar (\PT)}{W}}}{{\xi \PT + M{\Pc} + {\Ps}}}
\end{align}
where $\EEbar(\PT) = \frac{W{\SEbar(\PT)}}{{\xi \PT + M{\Pc} + \Ps}}$, and the derivative $\frac{{\de \SEbar \left( {\PT} \right)}}{{\de\PT}}$ is given by the optimal Lagrangian multiplier $\mu^*$ related to the power constraint of the SE maximization problem $\cP_5^{(\ell)}$. This concludes the proof of Property (ii) of \propref{theorem:properties_RE}.

\section{Proof of \propref{theorem:sumrate_KKT}}\label{app:D}
Note that $\cP_5^{(\ell)}$ is a convex program. Therefore, we can acquire its optimal solution ${\bLambda^{*}_{k}}$ through solving the corresponding KKT conditions. Note that $\frac{{\partial {{\cal L}}}}{{\partial \lambdak^* }}$ in \eqref{eq:solution_matrix} is a diagonal matrix. Then, the KKT conditions in \eqref{eq:solution_matrix} can be reduced to
\begin{align}\label{eq:KKT_app}
\left[ \frac{{\partial {\cL}}}{{\partial \lambdak ^{*}}} \right]_{m,m} & = {\frac{{\gamma _{k,m}^{*}}}{{1 + \gamma _{k,m}^{*}\lambda _{k,m}^{*}}}} - d ^{(\ell)}_{k,m} + \varphi _{k,m}^{*} - \mu ^* \ntb
&+ \sum\limits_{k' \ne k}^K {\sum\limits_{n = 1}^{N_{k'}} {\frac{{{{\widehat r}_{k',m,n}}}}{{\widetilde \gamma _{k',n}^{*} + {\sigma}^2 +  \tr {\bLambda_{\backslash k'}^{*}{{\widehat {\bf{R}} }_{k',n}}}}}} } \ntb
&  = 0,\quad m = 1,2, \ldots ,M
\end{align}
where $\varphi _{k,m}^{*}$ is the $m$th diagonal entry of ${\bf{\Psi }}_k^{*}$. Therefore, we can observe that the KKT conditions in \eqref{eq:KKT_1} and \eqref{eq:KKT_3} are equal to those of the following problem
\begin{align}\label{eq:ratemax_short}
\bLambda^{*} = \mathop {\arg \max }\limits_{\Lda} \quad & \sum\nolimits_{k} { {\log \det \left( {\bI_{M} + {{\bf{\Gamma }}_k}{\bLambda_k}} \right)} }\ntb
&{ + \log \det \left( {{{{\bf{\widetilde \Gamma }}}_k} + {{{\bf{\overline K}}}_k\left(\Lda\right)}} \right) - \tr { {{{\bf{D }}^{(\ell)}_{k}}{\bLambda_k}} }}\ntb
{\mathrm{s.t.}}\quad
& \sum\nolimits_{k=1}^{K}{\tr { \lambdak }} = \PT \ntb
& \lambdak \succeq \bzero,\; \lambdak\; \mathrm{diagonal},\; \forall k\in \K.
\end{align}
Note that \eqref{eq:ratemax_short} is also a convex program, whose KKT conditions are equivalent to those of $\cP_5^{(\ell)}$. Solving the corresponding KKT conditions, we have
\begin{align}\label{eq:Solution}
\left\{ \begin{array}{l}
\frac{{\gamma _{k,m}^{*}}}{{1 + \gamma _{k,m}^{*}\lambda _{k,m}^{*}}} + \sum\limits_{k' \ne k}^K {\sum\limits_{n = 1}^{N_{k'}} {\frac{{{{\widehat r}_{k',m,n}}}}{{\widetilde \gamma _{k',n}^{*} + {\sigma}^2 + \tr {{{\widehat {\bf{R}}}_{k',n}}\bLambda_{\backslash k'}^{*}}}}} }  \ntb
\qquad\qquad\qquad\qquad\qquad\qquad\quad\;\; = d ^{(\ell)}_{k,m} + {\mu ^{*}}, {\mu ^{*}} < \upsilon _{k,m}^{*}\ntb
\lambda _{k,m}^{*} = 0, \qquad\qquad\qquad\qquad\qquad\qquad\qquad\quad {\mu ^{*}} \ge \upsilon _{k,m}^{*}
\end{array} \right.
\end{align}
where the auxiliary variable $\upsilon _{k,m}^{*}$ is expressed as
\begin{align}
\upsilon _{k,m}^{*} & = \gamma _{k,m}^{*}  - d^{(\ell)} _{k,m} \ntb
+ &\sum\limits_{k' \ne k}^K {\sum\limits_{n = 1}^{N_{k'}} {\frac{{{{\widehat r}_{k',m,n}}}}{{\widetilde \gamma _{k',n}^{*} + {\sigma}^2 + \sum\limits_{\scriptstyle \ \ (l',m')\hfill\atop
\scriptstyle \in {\cal{S}}(k,m,k')\hfill} {{{\widehat r}_{k',m',n}}\lambda _{l',m'}^{*}} }}} }.
\end{align}
This concludes the proof.



\end{document}